\documentclass[a4paper,11pt]{article}
\pdfoutput=1 % if your are submitting a pdflatex (i.e. if you have
\usepackage{jcappub} % for details on the use of the package, please
\usepackage{bm}
\usepackage{url}
\usepackage{amsmath}
\usepackage{amsfonts}
\usepackage{morefloats}
\usepackage{natbib}
\usepackage{enumerate}
\usepackage{ulem}
\newcommand{\blue}{\color{blue}}

\newcommand{\sigmav}{\langle \sigma v \rangle}

\newcommand{\bbbar}{b \bar b}
\newcommand{\ww}{W^+W^-}

\title{Dark matter vs. astrophysics in the interpretation of AMS-02 electron and positron data}

\author[a,b]{Mattia Di Mauro}
\author[a,b]{Fiorenza Donato}
\author[a,b]{Nicolao Fornengo}
\author[a,b]{Andrea Vittino}

\affiliation[a]{Department of Physics, University of Torino, via P. Giuria 1, 10125 Torino, Italy}
\affiliation[b]{Istituto Nazionale di Fisica Nucleare, via P. Giuria 1, 10125 Torino, Italy}

\emailAdd{mattia.dimauro@to.infn.it}
\emailAdd{donato@to.infn.it}
\emailAdd{fornengo@to.infn.it}
\emailAdd{vittino@to.infn.it}

\abstract{We perform a detailed quantitative analysis of the recent AMS-02 electron and positron data. We investigate the interplay between the emission from primary astrophysical sources, namely Supernova Remnants and Pulsar Wind Nebulae, and the contribution from a dark matter
annihilation or decay signal. Our aim is to assess the information that can be derived on dark matter properties when both dark matter and primary astrophysical sources are assumed to jointly contribute to the leptonic observables measured by the AMS-02 experiment. We investigate both the possibility to set robust constraints on the dark matter annihilation/decay rate and the possibility to look for dark matter signals within realistic models that take into account the full complexity of the astrophysical background. Our results show that AMS-02 data enable to probe efficiently vast regions of the dark matter parameter space and, in some cases, to set constraints on the dark matter annihilation/decay rate that are comparable or even stronger than the ones derived from other indirect detection channels. 
}

\begin{document}
\maketitle
\flushbottom
\section{Introduction}

Dark matter (DM) stands out as one of the most intriguing challenges in physics. While gravitational effects of this elusive substance are manifest in a variety of observables, ranging from the astrophysical to the cosmological scale, an indisputable non-gravitational evidence of its existence has yet to be unveiled. Among the techniques developed to pursue the search of DM non-gravitational imprints, there is the so-called indirect detection, which consists in the attempt to reveal a signal that can be the outcome of a DM pair annihilation or decay event and that can possibly manifest its presence as an exotic component in cosmic rays (CRs).

In the framework of this searching strategy, a preeminent role is played by the analysis of the fraction of antimatter in the flux of charged CRs. In particular, in recent years, the observations performed by the PAMELA satellite of a steep rise in the energy spectrum of the positron fraction \cite{2009Natur.458..607A} has been interpreted by many authors as a possible evidence for the existence of a leptophilic TeV-scale WIMP (see, for example, Refs.~\cite{Bergstrom:2008gr,Cholis:2008hb,Cirelli:2008pk,Bertone:2008xr,Nardi:2008ix,Hooper:2008kv,Hooper:2009fj,Dev:2013hka}).  However, as stressed in several works,  plausible astrophysical mechanisms able to generate such a rise in the positron fraction can be conceived: while the most widely known example invokes Pulsar Wind Nebulae (PWNe) or Supernova Remnants (SNRs) or a combination of both as sources of primary $e^\pm$ \cite{Hooper:2008kg,Yuksel:2008rf,Profumo:2008ms,Fujita:2009wk,Kohri:2015mga}, it has also been shown that the secondary production of $e^{\pm}$ {\it within} the shock region of SNRs provides a perfectly viable mechanism to explain observations \cite{Blasi:2009hv,PhysRevD.80.123017}. Several features that characterize these interpretations of the PAMELA excess have been investigated in a series of recent works (see, for example, Refs.~\cite{Lavalle:2014kca,Boudaud:2014dta,Mertsch:2014poa,Lin:2014vja,Jin:2014ica,Delahaye:2014osa,Gaggero:2013nfa,Ibarra:2013zia,Bergstrom:2013jra,Li:2014csu,2014PhLB..728..250F}) in connection to the data delivered by the AMS-02 experiment. 

In a previous work \cite{DiMauro:2014iia} we have carried out a detailed and quantitative analysis of the four observables measured by the AMS-02 experiment (electron flux, positron flux, total leptonic flux and positron fraction) within a purely astrophysical model for $e^{\pm}$ generation. In this paper our purpose is to extend this quantitative study to include also a DM contribution: the aim is to assess the constraints on DM obtainable within a model in which the astrophysical background is included and modeled in a realistic way. In particular, we stress that, differently from Ref.~\cite{DiMauro:2014iia} where we used an 8\% error on the AMS-02 measured fluxes, we are now making full profit of the recently published  AMS-02 Collaboration data, which report smaller uncertainties on their whole set of observables.

The paper is organised as follows: in Section \ref{sec:contributions} we briefly illustrate the various contributions to the electron and positron fluxes, while in Section \ref{sec:transport} we describe the framework that we use to model the transport of positrons and electrons across the Galaxy and the heliosphere. In Section \ref{sec:results}, we present the methods that we use for the different parts of our analysis and the results that we obtain. Finally, in Section \ref{sec:conclusions} we derive our conclusions. 

%%%%%%%%%%%%%%%%%%%%%%%%%%%%%%%%
%%%%%%%%%%%%%%%%%%%%%%%%%%%%%%%%
%%%%%%%%%%%%%%%%%%%%%%%%%%%%%%%%
\section{Contributions to the $e^+e^-$ flux}
\label{sec:contributions}
In this Section, we briefly review the role of the main sources that contribute to the four observables measured by the AMS-02 experiment. For a more detailed description of all the astrophysical contributions (primary and secondary) we address the reader to Ref.~\cite{DiMauro:2014iia}. 
\subsection{Secondaries}
\label{subsec:secondaries}
Electrons and positrons can be generated in spallation reactions that involve primary CRs, in particular proton and Helium nuclei, impinging on the nuclei that populate the interstellar medium (ISM), the dominant component of which being represented by Hydrogen. The source term related to this contribution is: 
\begin{eqnarray}
\label{eq:source}
  q_{e^{\pm}}(\vec{x},E_{e}) = 4 \pi \; n_{\rm ISM}(\vec{x}) \displaystyle 
\int dE_{\rm CR} \Phi_{\rm CR} \left( \vec{x} , E_{\rm CR} \right) \frac{d\sigma}{dE_{e}}(E_{\rm CR}, E_{e}) ,
\end{eqnarray}
where the quantity $n_{\rm ISM}$ represents the density of target nuclei in the ISM, $\Phi_{\rm CR}$ is the flux of the primary CR species and the term $d\sigma / dE_{e}$ stands for the differential inclusive cross section for the electron/positron production in the spallation reaction under study. 

\begin{figure}
\centering
\includegraphics[width=0.7\textwidth]{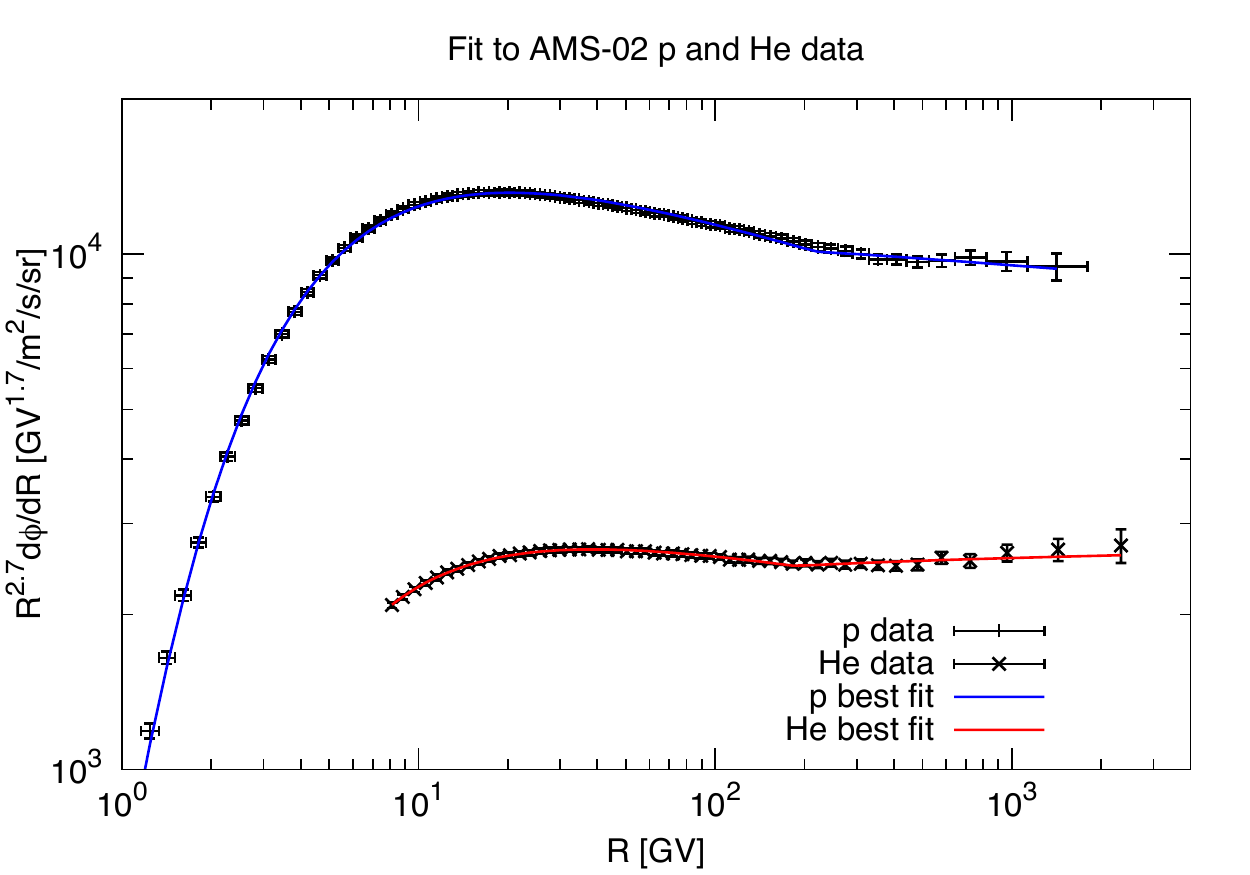}
\caption{Fit to the proton and helium fluxes measured by AMS-02: the black points denote experimental data, while the blue and red solid lines represent our best fits.}
\label{fig:pHe_ams02}
\end{figure}

By following the same approach outlined in Ref.~\cite{DiMauro:2014iia}, we set the primary CR fluxes by fitting the recent AMS-02 measurements of the proton \cite{Aguilar:2015ooa} and Helium \cite{AMS-HE} 
fluxes with the following interstellar spectra: 
\begin{eqnarray}
\label{eq:conddm}
\Phi_{\rm CR}(R) =  
\left\{
\begin{array}{rl}
& A \beta^{P}R^{-P_1} \rm{\;for\;} R \leq R_{\rm break}, \\
& A \beta^{P}R_{\rm break}^{-P_1+P_2}R^{-P_2}  \rm{\;for\;} R > R_{\rm break},
\end{array}
\right.
\label{eq:phi}
\end{eqnarray}
where $R=pc/eZ$ and $\beta$ denote, respectively, the rigidity and the velocity of the particle under consideration. We allow for the presence of a spectral break at a rigidity $R_{\rm break}$ to be determined through the fitting procedure. For the best-fit configuration the values of the parameters are: $A=26700 \pm  500$ m$^{-2}$s$^{-1}$sr$^{-1}$(GeV/n)$^{-1}$, $P=7.2 \pm 0.4$, $P_1=2.877 \pm 0.004$, $P_2=2.748 \pm 0.013$ and $R_{\rm break} = 220 \pm 22$  GV for the proton and $A=4110 \pm  80$ m$^{-2}$s$^{-1}$sr$^{-1}$(GeV/n)$^{-1}$, $P=3.5 \pm 0.7$, $P_1=2.793 \pm 0.004$, $P_2=2.689 \pm 0.013$ and $R_{\rm break} = 187 \pm 18$ GV for the Helium, the Fisk solar modulation potential being set to 700 MV. 
The best fit to the combined proton and Helium data has a reduced chi-square ($\chi^2$/d.o.f.) equal to 0.43.% which corresponds to a $p$-value and significance of $3.1\times10^{-9}$ and $5.9\sigma$, respectively.
The proton and Helium fluxes for this best-fit configuration are shown, together with the AMS-02 measurements in Fig.~\ref{fig:pHe_ams02}.    

Concerning the inclusive production cross sections $d\sigma / dE_{e}$,  for the case of p-p collisions we use the parameterization described in Ref.~\cite{2006ApJ...647..692K}, while to model processes involving Helium nuclei, either as the incoming particle or the target, we resort to the empirical prescriptions illustrated in Ref.~\cite{2007NIMPB.254..187N}. 
\subsection{Supernova Remnants}
\label{subsec:SNRs}
SNRs are usually thought to play the dominant role in accelerating charged cosmic rays within the Galaxy. This acceleration is realised through the propagation of non-relativistic shock waves that are produced by the star explosion. The typical SNR injected spectra have the shape of a power-law with a cut-off at large energies: 
\begin{equation}
     \label{Q}
         Q(E)=Q_{0, \rm SNRs} \left(\frac{E}{E_0}\right)^{-\gamma_{\rm SNRs}}\exp{\left(-\frac{E}{E_c}\right)},
\end{equation} 
being $Q_{0, \rm SNRs}$ the normalization of the spectrum, $\gamma_{\rm SNRs}$ the spectral index and $E_c$ the cut-off energy. As it has been extensively discussed in \cite{DiMauro:2014iia}, the parameters $Q_{0, \rm SNRs}$ and $\gamma_{\rm SNRs}$ can be inferred from the study of the radio emission measured in the region of the sky that hosts the SNR, while $E_c$ is expected to lie at the TeV range \cite{2008PhRvL.101z1104A,2009ApJ...692.1500A,2010ApJ...714..163A,Aharonian:2001mz,Aharonian:2006ws}. In the analysis that will be described in the following Sections of the paper, we assume $E_c = 2$ TeV. Let us remark that, however, as far as $E_c$ is beyond the maximal energy that is measured by AMS-02, the exact value of this parameter does not affect our results. Finally, every time that, in our investigations, we will need to use SNRs parameters we will resort to the Green catalogue \cite{Green:2014cea}, which is the most complete catalogue of Galactic SNRs.    

\subsection{Pulsar Wind Nebulae}
\label{subsec:PWNe}
Also pulsars are expected to generate a flux of electrons and positrons through a mechanism known as spin-down emission (for a description of which we address the reader to Refs.~\cite{1970ApJ...162L.181S,1987ICRC....2...92H,1996SSRv...75..235A,2001AA...368.1063Z,Amato:2013fua,2013arXiv1312.3483D,Ruderman:1975ju,1976ApJ...203..209C,Cheng:1986qt}). In a few simple words, this mechanism is a consequence of electrons and positrons being torn away from the surface of the neutron star by the strong electric field genarated by the pulsar spinning. These charged particles gather in a sort of wind that surrounds the pulsar and then are released in the ISM at the disruption of this nebula. Because of this injection mechanism, which is fast and followed by a weak residual energy emission, PWNe can be considered as burst-like sources of $e^\pm$.  

The spectrum of the $e^\pm$ injected by a PWN in the ISM has the same expression as the one in Eq.~\ref{Q} associated to SNRs. As outlined in \cite{DiMauro:2014iia}, the normalization of this spectrum is related to the efficiency $\eta_{\rm PWNe}$ with which the PWN can convert its spin down energy into the production of $e^\pm$ pairs: 
\begin{equation}
\label{Wop}
\int_{E_{\rm min}}^{\infty}\,dE\,E\,Q(E)\,=\eta_{\rm PWNe}\,W_0,
\end{equation}
where the quantity $W_0$ represents the total spin-down energy which, in terms of the present age of the pulsar $t_*$ and the typical pulsar decay time $\tau_0$ can be expressed as:
\begin{equation}
W_0\approx\tau_0\dot{E}\left(1+\frac{t_*}{\tau_0}\right)^2.
\end{equation}

The most complete list of PWNe is represented by the ATNF catalogue \cite{1993ApJS...88..529T}. As it will be widely discussed in the following, we will use it as a reference for all the PWN parameters.

\subsection{Dark Matter}
\label{subsec:DM}
Positrons and electrons can also be the result of the pair annihilation or decay of DM particles. The source terms associated to these contributions are:
\begin{equation}
\begin{aligned}
{\cal Q}_{\rm ann}(\vec{x},E) &=& \epsilon \left(\frac{\rho(\vec{x})}{m_{DM}}\right)^2\sum_f \langle \sigma v \rangle_f \frac{dN^f_{e^{\pm}}}{dE}, \\
{\cal Q}_{\rm dec} (\vec{x},E) &=& \left(\frac{\rho(\vec{x})}{m_{DM}}\right)\sum_f \Gamma_f \frac{dN^f_{e^{\pm}}}{dE},~~~~~
\end{aligned}
\label{eq:Q_dm}
\end{equation}
where $\vec{x}$ denotes Galactic position, 
$\epsilon$ being a factor that takes the value 1/2 or 1/4 for, respectively, a self-conjugate or non self-conjugate DM particle, while $f$ denotes the Standard Model particles that can be produced in the annihilation or decay process and the functions $dN^f_{e^{\pm}}/dE$ represent the $e^{\pm}$ energy spectrum generated in a single annihilation or decay process. 
The galactic DM halo, filled with particles with mass $m_{DM}$, follows  a spatial density $\rho(\vec{x})$.
We perform a model independent analysis which consists in assuming that the DM annihilation/decay occurs in a single channel. In particular, we will focus our attention on the five channels $e^+e^-$, $\mu^+\mu^-$, $\tau^+\tau^-$, $\bbbar$, $\ww$. We model the energy spectra $dN^f_{e^{\pm}}/dE$ from Ref.~\cite{Cirelli:2010xx}: we remind that these spectra have been computed by taking into account {\it electroweak correction} which, as stressed in \cite{Ciafaloni:2010ti} can play a non-negligible role in shaping the $e^{\pm}$ emission when the DM mass is above the electroweak scale.

%%%%%%%%%%%%%%%%%%%%%%%%%%%%%%%%
%%%%%%%%%%%%%%%%%%%%%%%%%%%%%%%%
%%%%%%%%%%%%%%%%%%%%%%%%%%%%%%%%
\section{Transport in the galaxy and in the Heliosphere}
\label{sec:transport}
After being injected by their source, electrons and positrons propagate across the ISM where spatial diffusion, convection, reacceleration and energy losses can shape their spectrum. To model this transport and take into account these effects, we have to solve the transport equation:
\begin{equation}
\partial_t {\cal N} - \vec{\nabla}\cdot 
\left\{ K(E)  \vec{\nabla}{\cal N} \right\} + 
\partial_E \left\{ \frac{dE}{dt} {\cal N} \right\} = {\cal Q}(E,\vec{x},t)\;,
\label{eq:prop}
\end{equation}
where ${\cal N}$ denotes the positron(electron) number density per unit energy while ${\cal Q}(E,\vec{x},t)$ represents the source terms that were 
discussed in the previous Sections. The diffusion coefficient $K(E)$, which we assume to be space independent, is defined in the following way: 
\begin{equation}
K(E)= \beta K_0 ( R/{1~\rm GV})^{\delta}.
\end{equation}
The  term $\frac{dE}{dt}$ is given by the sum of the synchrotron energy loss term (which depends on the value of the Galactic magnetic field, for which, compatibly with Ref.~\cite{Sun:2007mx}, we have assumed an intensity of 2 $\mu$G (3 $\mu$G) for the regular (random) component) and the inverse Compton (IC) loss term (for which we have included a full relativistic treatment \cite{2010AA...524A..51D}). Additional energy loss processes, namely bremsstrahlung, ionization and Coulombian interactions on the ISM can be safely neglected.  

In solving Eq.~\ref{eq:prop}, we adopt the  \emph{two-zone diffusion model}, which has been extensively described in literature (see for example Ref~\cite{2001ApJ...555..585M}). 
In this paper we adopt the set of propagation parameters defined in the {\sc Med} propagation model of Ref.~\cite{2004PhRvD..69f3501D}.  For any further detail  we address the reader to Ref.~\cite{2010AA...524A..51D}. 

Lastly, for the solar modulation of the fluxes, we have used the usual force field approximation treated in Refs.~\cite{1971JGR....76..221F, 1987AA...184..119P}. In this frameworks, the solar modulation depends on the value of the parameter $\phi$, the so-called {\it Fisk potential}. As it will be discussed in the following, we will always determine $\phi$ directly from data.  

%%%%%%%%%%%%%%%%%%%%%%%%%%%%%%%%
%%%%%%%%%%%%%%%%%%%%%%%%%%%%%%%%
%%%%%%%%%%%%%%%%%%%%%%%%%%%%%%%%
\section{Analysis and results}
\label{sec:results}
The AMS-02 Collaboration has recently published the results of their 30 months measurements of the full set of observables that are related to the physics of cosmic electrons and positrons, {\it i.e.}, the $e^-$ and $e^+$ fluxes \cite{2014PhRvL.113l1102A}, the total $e^- + e^+$ flux \cite{2014PhRvL.113v1102A} and the positron fraction $e^+/(e^- + e^+)$ \cite{2014PhRvL.113l1101A}. Our main purpose in this Section is to test various possible interpretations of these experimental data.  

The analyses are based on a global fit of the four aforementioned observables within several theoretical frameworks (described below), each one being characterized by a certain set of parameters $\{ \theta_1, \dots \theta_N \}$, whose total number $N$ depend on the specific model under scrutiny. The $N$-dimensional parameters space of each model is sampled by means of a Markov Chain Monte Carlo (MCMC) scan performed with the COSMOMC package \cite{Lewis:2002ah}. For every parameter considered in our analyses, a flat prior is assumed. In order to determine the configuration of parameters providing the best agreement with AMS-02 measurements we assign to each N-tuple $\{ \theta_1, \dots \theta_N \}$ a chi-square estimator defined as:
\begin{equation}
\chi^2 = - 2{\rm ln} L = \sum_j\sum_i\frac{(f^j_i(\theta_1, \dots \theta_N )-d^j_i)^2}{({\sigma_i^j})^2},
\end{equation}   
where the indices $j$ and $i$ run over the four AMS-02 observables and the different energy bins,  respectively. The quantities $f_i^j(\theta_1, \dots \theta_N )$ represent the theoretical predictions while $d_i^j$ stands for the observed experimental data values, which are affected by the experimental uncertainties $\sigma_i^j$.  

The experimental uncertainties $\sigma_i^j$ are defined as the sum in quadrature of the statistical and systematic uncertainties. The latter, reported in \cite{2014PhRvL.113l1102A,2014PhRvL.113v1102A,2014PhRvL.113l1101A}, are associated to {\it event selection}, determination of the {\it acceptance}, {\it bin-to-bin migration} and {\it charge confusion} (that, obviously, affects all the observables except the total $e^+ + e^-$ flux) and in the interval of energies that we consider in our analysis have an impact that ranges from a few \% (for energies around 10 GeV) to 10-15 \% (for the highest energies).

%As already done in Ref.~\cite{DiMauro:2014iia}, we choose to test the different theoretical models by analyzing the entire set of observables measured by AMS-02: electron flux, positron flux, total $e^+ + e^-$ flux and positron fraction. In fact, even if, under a purely formal point of view, only two of them should be considered as truly independent, this is not the case in practice, since the four observables are the result of independent experimental analyses. Therefore, we believe that using the four datasets is a more robust way to check the consistency among the measurements and their agreement with the theoretical models that we consider.    
Another point worth to be mentioned here is that in this analysis we choose to consider only data above 10 GeV.  In fact, the sector relevant for the investigation of primary astrophysical $e^\pm$ sources and of  the corresponding DM contribution is the high energy part of the spectra. Considering that time-dependent solar modulation effects become increasingly important at lower energies, we decide to concentrate our analysis only on the high energy part, in order to avoid that the numerous AMS-02 energy bins at very low energy (which require detailed, although uncertain, solar modulation modeling) would drive the statistical significance of the fit to the part of the spectrum which is of least importance to the present analysis. Since the effect of transport in the heliosphere is still present up to few tens of GeV, we nevertheless consider solar modulation by applying the simple Fisk method (as discussed above): however, by concentrating on energies above 10 GeV we minimize the bias of the low-energy solar-modulation-dominated data bins. The choice of 10 GeV as the discriminating energy has been adopted as a trade off between the requirement of reducing the impact of solar modulation modeling and the interest to investigate the energy region where DM and primary sources start to emerge. 

%%%%%%%%%%%%%%%%%%%%%%%%%%
\subsection{The purely astrophysical interpretation}

In this paragraph, we follow the prescriptions presented in our previous work \cite{DiMauro:2014iia} and discuss our new analysis for a purely astrophysical interpretation of AMS-02 measurements. In brief, the model, which we label {\it astro model}, is composed by: 
\begin{itemize}
\item{{\bf Secondary $e^-$ and $e^+$}: their flux is computed by using the source term introduced in Section \ref{subsec:secondaries}. We recall that this component is fixed. In fact, its contribution depends on the primary CR fluxes and on the parameterization of the spallation cross sections, which we do not vary in our  analysis. The secondary $e^+$ components is the dominant one at low energies, where it has been shown in Ref.~\cite{DiMauro:2014iia} to perfectly reproduce the measured AMS-02 flux, within the chosen propagation model.}
\item{{\bf Primary $e^-$ accelerated by SNRs}: for this contribution, which has been discussed in Section \ref{subsec:SNRs}, we follow the approach outlined in Ref.~\cite{DiMauro:2014iia}, that consists in separating the total population of SNRs into two categories: 
\begin{itemize}
\item{Far SNRs ($d > 3$ kpc): they are modeled as an average component with a spatial distribution  taken from Ref.~\cite{2004IAUS..218..105L}. The sources that fall in this class share common values for the normalization $Q_{0,{\rm SNRs}}$ and the spectral index $\gamma_{\rm SNRs}$ of the 
injected electron flux. These parameters are left {\it free} to vary in our fitting procedure.}
\item{Near SNRs ($d < 3$ kpc): the parameters that characterize these sources,  {\it i.e.} their distances, ages, spectral indices and the normalization of their fluxes, are taken directly from the Green catalogue \cite{Green:2014cea}, as anticipated in Section \ref{subsec:SNRs} and extensively discussed in Ref.~\cite{DiMauro:2014iia}. Notice that, in order to allow for a better agreement with the electron flux at high energies, we introduce a free normalization for the Vela SNR, $N_{\rm Vela}$,  which therefore represents an additional parameter, not included in our previous analysis of Ref.~\cite{DiMauro:2014iia}\footnote{We keep the distance and age of Vela fixed to, respectively, $d_{\rm Vela}=0.294$ kpc and $T_{\rm Vela}=11.3$ kyr}. Freedom in the variation of this parameter can be ascribed to variations in the value of the magnetic field of the source: $N_{\rm Vela} \propto B^{-(\gamma_{\mathrm{SNRs}}+1)/2}$, $N_{\rm Vela} = 1$ corresponding to $B=30$ $\mu$G.}
\end{itemize}}
\item{{\bf Primary $e^-$ and $e^+$ accelerated by PWNe}: by using the same approach of Ref.~\cite{DiMauro:2014iia}, we consider the PWNe of the ATNF calatog \cite{1993ApJS...88..529T} as primary sources. The pulsars spin-down energies are derived from the catalog itself. The 
 efficiency $\eta_{\rm PWNe}$ of conversion of this energy into electrons and positrons, and the spectral index $\gamma_{\rm PWNe}$  of the emission are assumed to be common to all the pulsars in the catalog and are free to vary.}
\end{itemize} 

The contributions to the $e^{\pm}$ fluxes coming from all these sources are propagated in the ISM and modulated in the heliosphere according to the prescriptions discussed in Section \ref{sec:transport}.  While the reference model for the Galactic propagation is described by the {\sc Med} set of parameters,  the Fisk potential $\phi$ of the solar modulation is a free parameter. To summarize, the {\it astro model} is characterized by six free parameters: $Q_{0,{\rm SNRs}}$, $\gamma_{\rm SNRs}$, $N_{\rm Vela}$, $\eta_{\rm PWNe}$, $\gamma_{\rm PWNe}$ and $\phi$.  

\begin{figure}[t]
\centering
\includegraphics[scale =0.5]{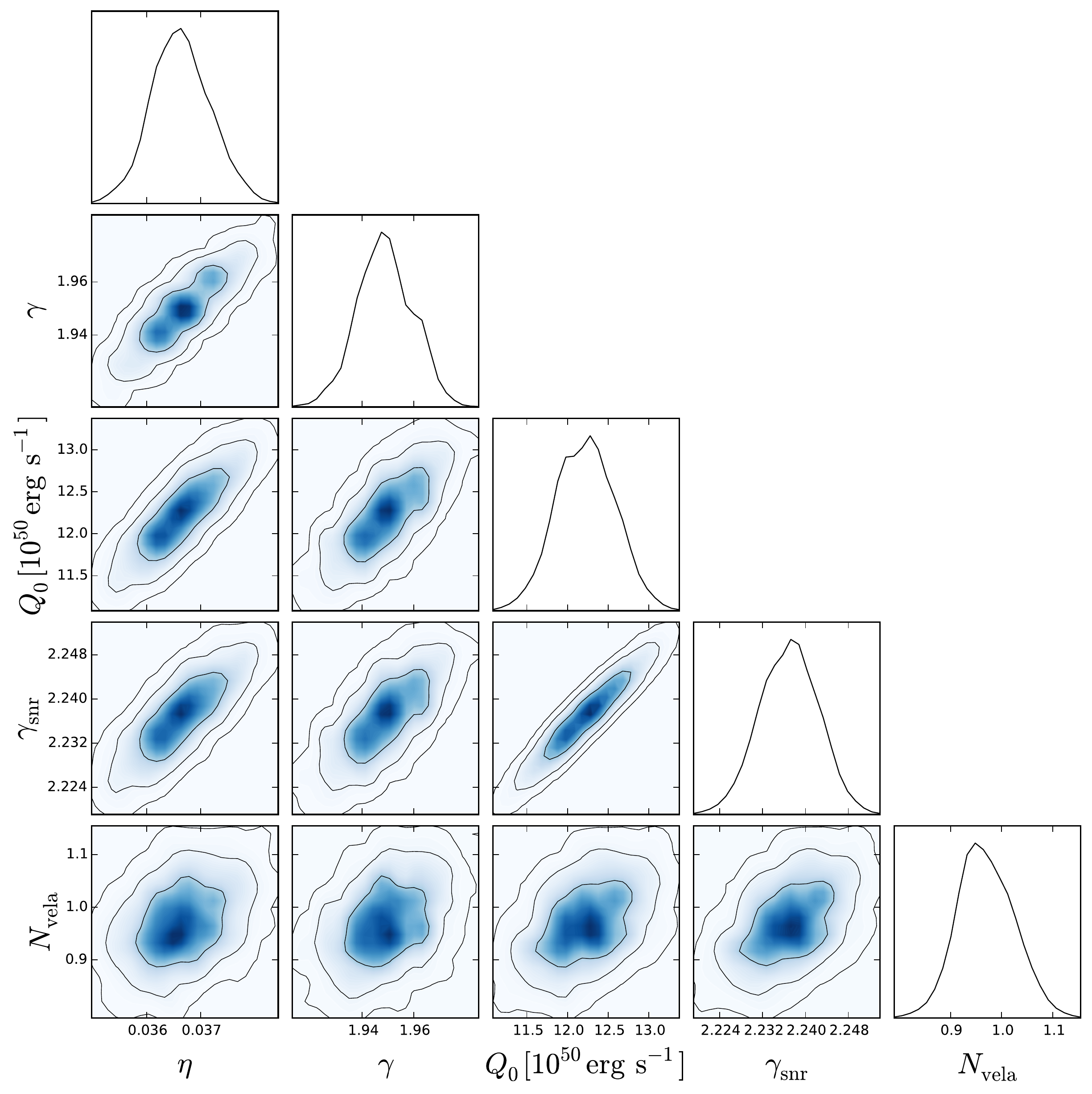}
\caption{Triangular plot for the fit to the four AMS-02 datasets of the {\it astro model} parameters, as reported in the first column of Table \ref{tab:astro}. The contours refer to $1\sigma$, $2\sigma$ and $3\sigma$ C.L. allowed regions. The plots along the diagonal show the posterior distribution for each parameter.}
\label{fig:astro_tri}
\end{figure}

 We start our analysis of the {\it astro model} by fitting the whole amount of data released by AMS, which consists of the four datasets corresponding to the positron flux, the electron flux, the total $e^+$ + $e^-$ flux and the positron fraction. Fig. \ref{fig:astro_tri} illustrates the posterior probability distributions of the parameters of the model, as they are sampled by the MCMC scan. Their values for the best-fit configuration are reported in the first column of Table \ref{tab:astro}, together with the global chi-square and the chi-square associated to each dataset. While the overall fit is good, the contribution to the total chi-square that comes from the positron fraction is large (the $\chi^2$ per data point is close to 2). This points toward a hint that the {\it astro} model might have some issues in reproducing the behaviour of this observable.
 
\begin{figure}[t]
\centering
\includegraphics[width=0.49\textwidth]{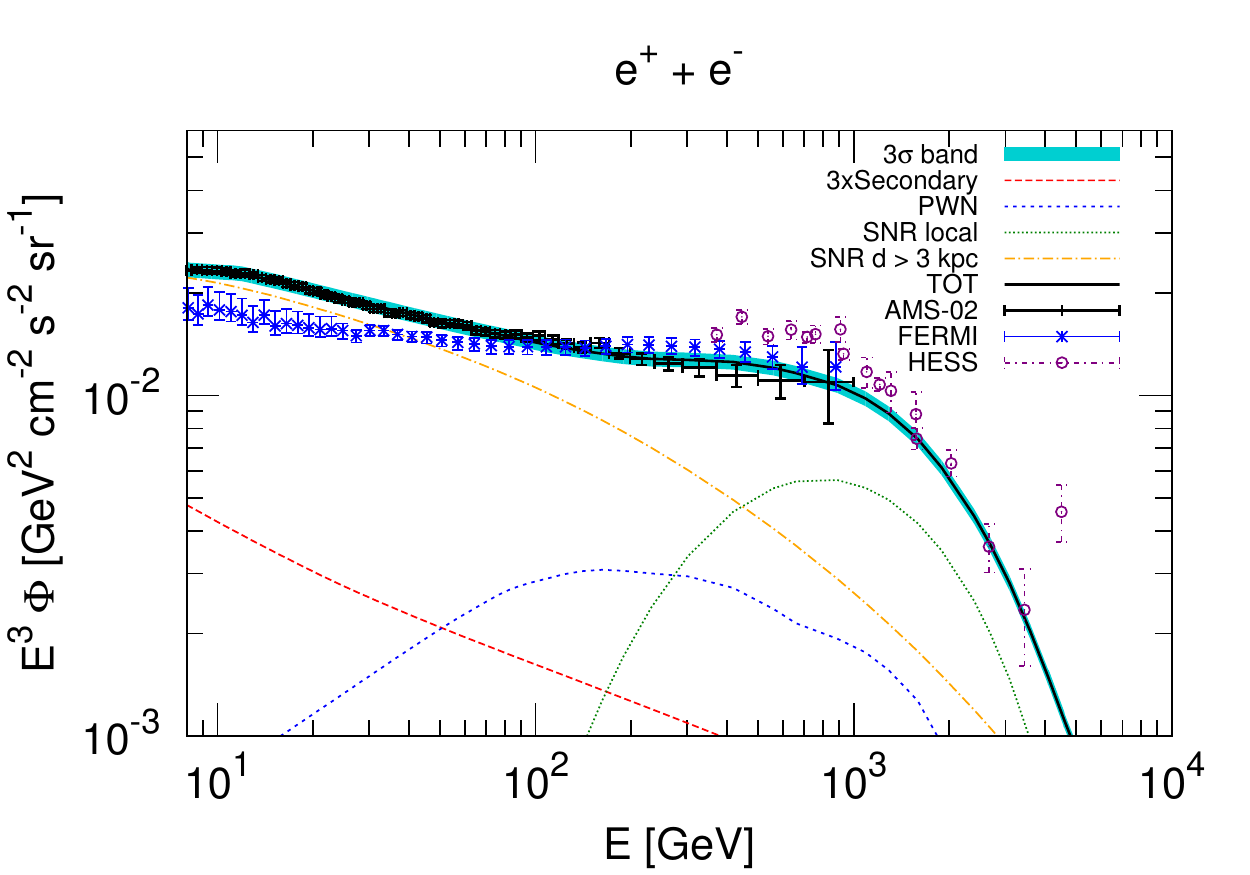}
\includegraphics[width=0.49\textwidth]{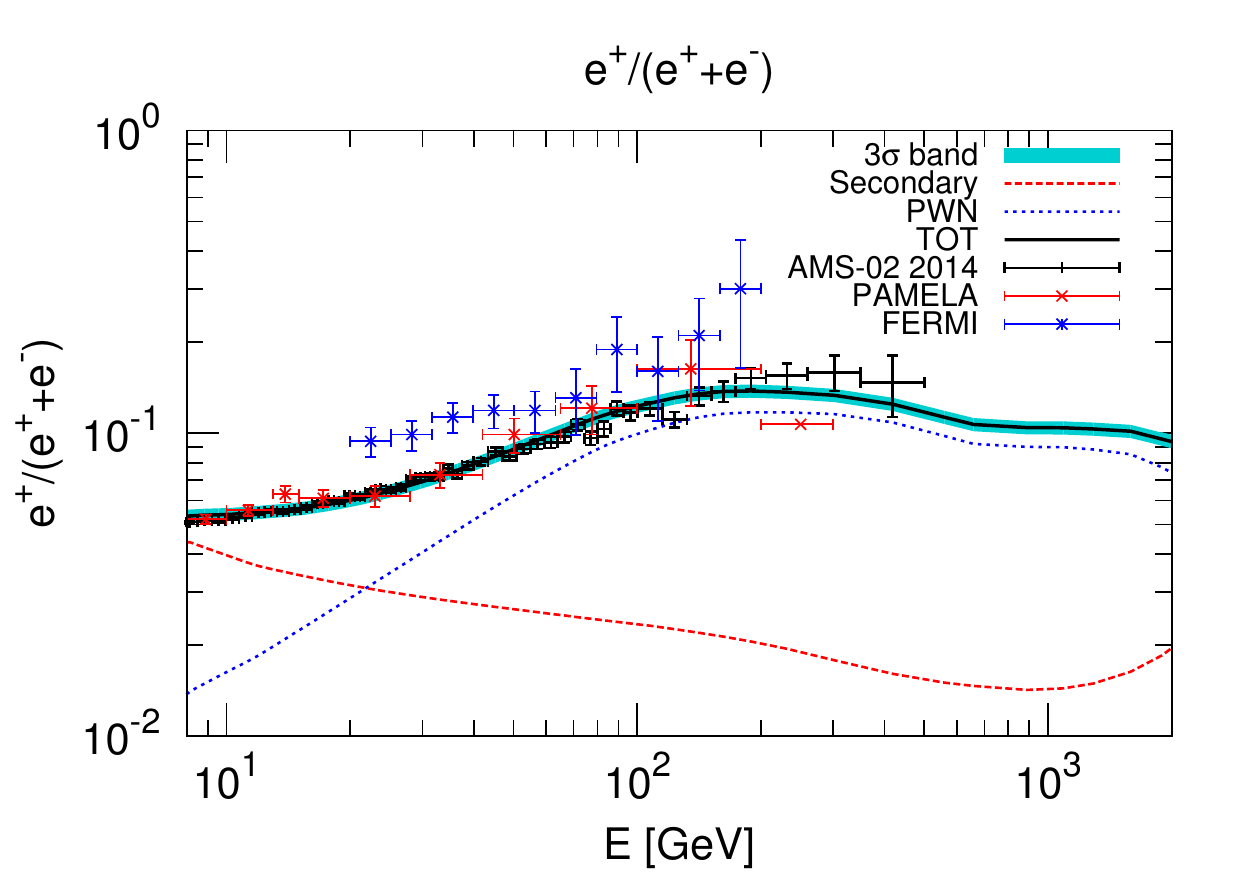}
\caption{Astrophysical fit ({\it astro model}) to the AMS-02 positron fraction and total flux datasets \cite{2014PhRvL.113l1102A,2014PhRvL.113v1102A,2015PhRvL.114q1103A}. The fit refers to the second column of Table \ref{tab:astro}. 
We display also Fermi-LAT \cite{2012PhRvL.108a1103A,2010PhRvD..82i2004A}, 
PAMELA \cite{2009Natur.458..607A,2011PhRvL.106t1101A,2013arXiv1308.0133P}, 
and HESS data \cite{2008PhRvL.101z1104A,2009AA...508..561A}.
The styles and colors used to represent the various contributions are described in the insets. }
\label{fig:astro_fit}
\end{figure}

\begin{figure}
\centering
\includegraphics[width=0.49\textwidth]{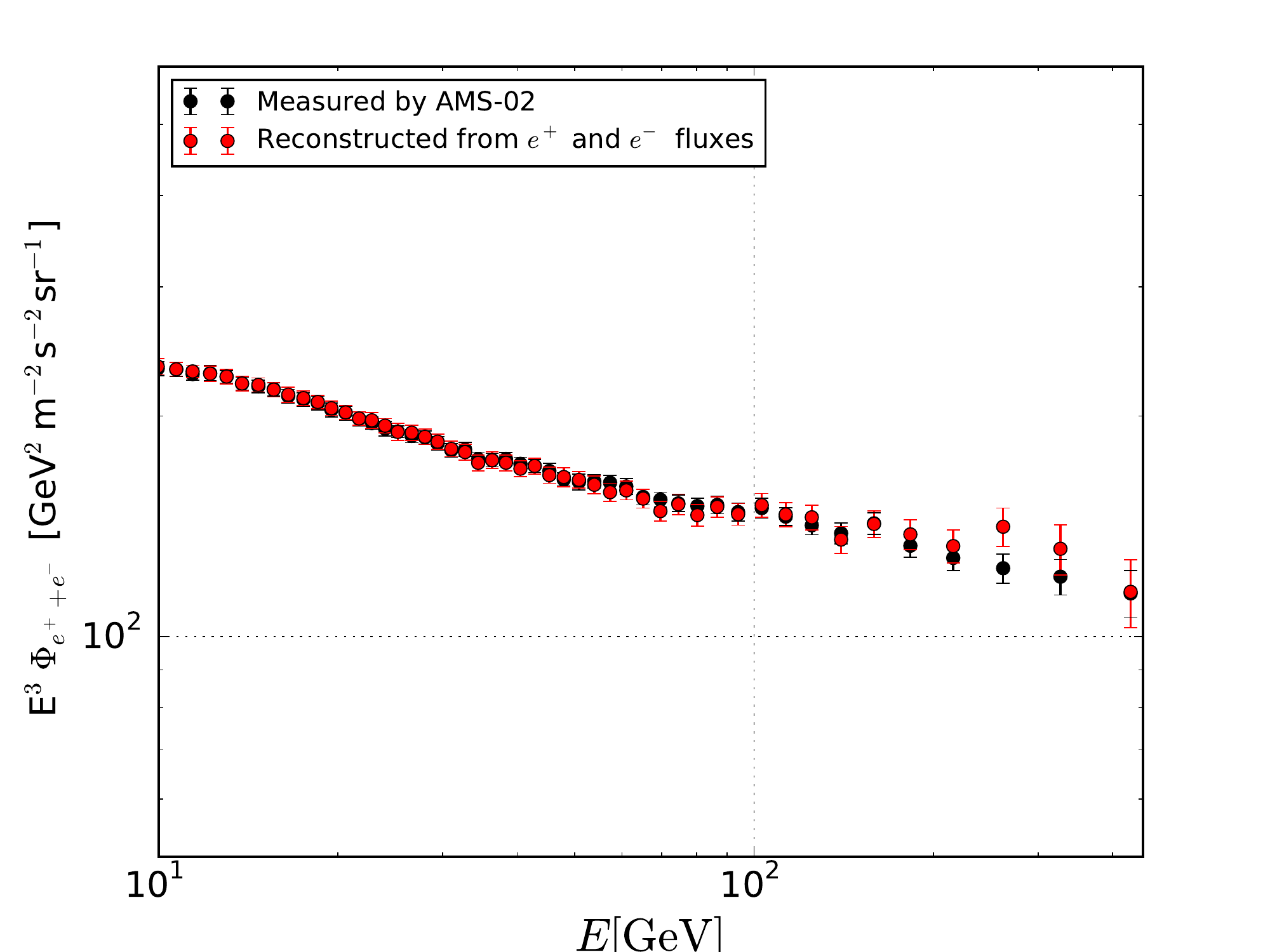}
\includegraphics[width=0.49\textwidth]{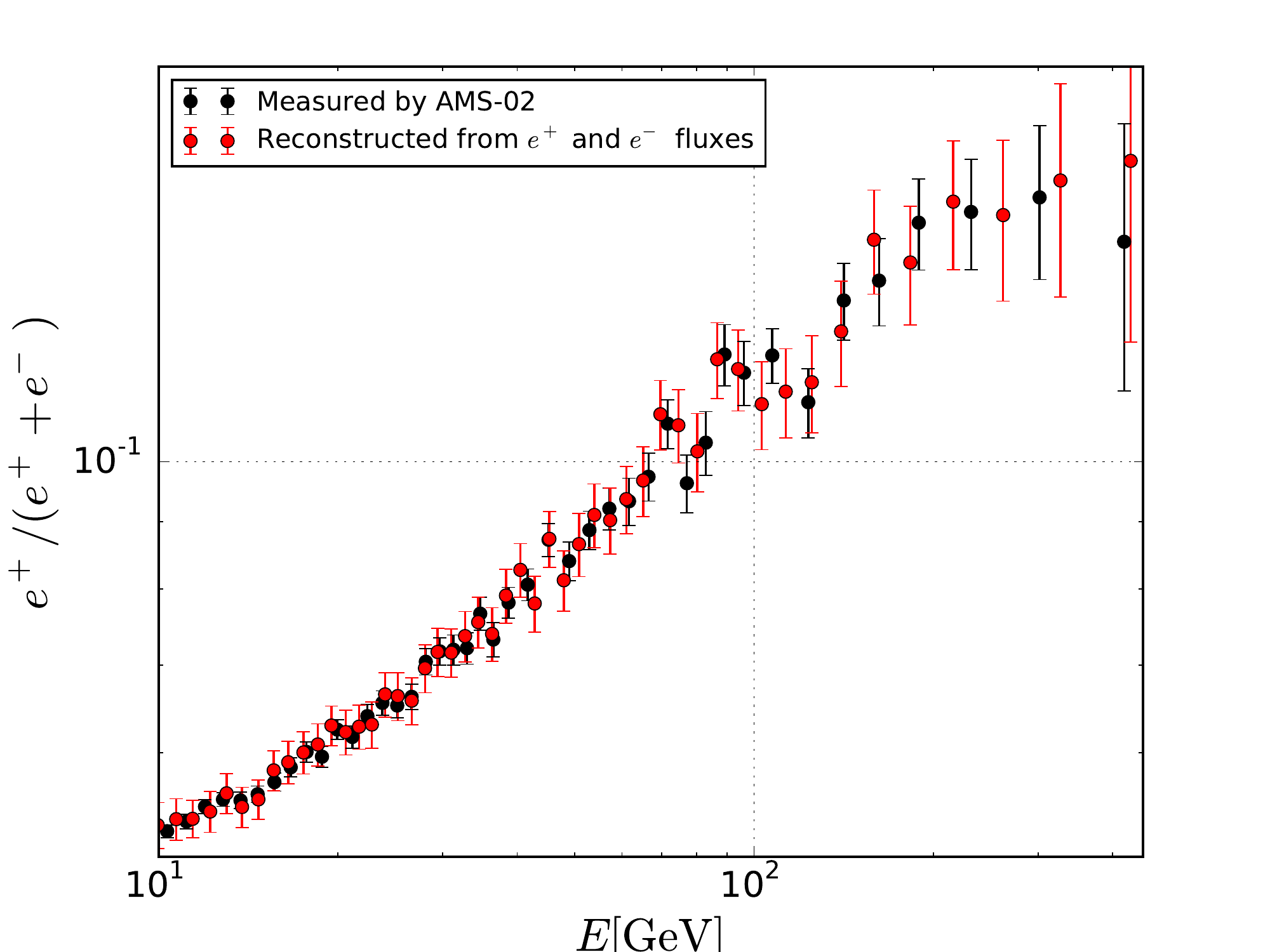}
\caption{Comparison between the directly measured total flux and positron fraction (black points) and the corresponding observables reconstructed from the measured single positron and electron fluxes (red points).}
\label{fig:compatibility}
\label{fig:AMS-comp}
\end{figure}

To have a clearer view about this fact, we separate the four datasets in two independent groups, and fit them separately: in one case we use the single electron and positron fluxes, in the other case the positron fraction and the total electron+positron flux. The second column of Table \ref{tab:astro} reports the best-fit parameters obtained by fitting only the positron fraction and the total flux, while the third column illustrates the outcome of a fit performed only on the positron and electron fluxes. The best-fit configurations for the former case are shown, together with AMS-02 data, in Fig. \ref{fig:astro_fit}. The results confirm that the {\it astro} model has no problem in predicting single and total fluxes that are in a agreement with the ones measured by AMS-02, but it has some tension with the measured positron fraction\footnote{We have checked that analogous results are obtained if different combinations of datasets are considered. }.  Notice that the AMS-02 data are internally perfectly consistent, as can be seen in Fig. \ref{fig:AMS-comp} where the positron fraction and total flux derived from the single fluxes are plotted together with the corresponding measured quantities. We have explicitly checked that the deviation between the reconstructed and measured quantities is always well within the uncertainty of the reconstructed values, which is determined by adopting the usual error propagation prescriptions (the only exception being the total flux at 262 GeV which however does not represent an issue at all in our analysis). The main difference arises from the much smaller experimental uncertainty on the directly-measured position fraction (especially for energies below 30 GeV), as a consequence of systematic effects that cancel out in the determination of ratios.

\begin{table}[t]
\center
\begin{tabular}{|c|c|c|c|}
\hline
parameter & all 4 datasets & only PF and SUM & only $e^+$ and $e^-$ fluxes \\
\hline 
${\eta_{\rm PWNe}}$& $0.037\substack{+0.001 \\ -0.001} $ & $0.036\substack{+0.002 \\ -0.001}$ & $0.037\substack{+0.001 \\ -0.002}$ \\
\hline 
$\gamma_{\rm PWNe}$& $1.95\substack{+0.03 \\ -0.02} $ & $1.94\substack{+0.04 \\ -0.02} $ & $1.95\substack{+0.2 \\ -0.2} $\\
\hline
$Q_{0, \rm SNRs}[10^{50}$ erg/s]   & $1.23\substack{+0.01\\ -0.03} $ & $1.10\substack{+0.15 \\ -0.05}$ & $1.26\substack{+0.06 \\ -0.09}$\\
\hline
$\gamma_{\rm SNRs}$ & $2.24\substack{+0.02 \\ -0.01} $ & $2.22 \substack{+0.02 \\ - 0.01}$  & $2.24 \substack{+0.01 \\ - 0.01}$\\
\hline
$N_{\rm Vela}$ &$0.98\substack{+0.03 \\ -0.13} $ & $1.00 \substack{+0.23\\ -0.19} $  &$0.93 \substack{+0.14 \\ -0.16} $ \\
\hline
$\chi^2_{\mathrm{tot}}/\mathrm{d.o.f}$ & 1.03 & 1.35 &  0.76\\
\hline
$\chi^2_{\mathrm{pf}} $ (43 data pts) & 81.7 & 80.4& - \\
\hline
$\chi^2_{\mathrm{sum}}$ (50 data pts) & 36.0 &  37.2 & - \\
\hline
$\chi^2_{e^+}$ (49 data pts) & 39.7 & - & 40.4 \\
\hline
$\chi^2_{e^-}$ (49 data pts)  & 33.6 &-  &28.9 \\
\hline

\end{tabular}
\caption{Best-fit parameters for the {\it astro model}, together with the $\chi^2$ associated to the different datasets,
for different combinations of the datasets.}
\label{tab:astro}
\end{table}

\begin{figure}[t]
\centering
\includegraphics[width=0.49\textwidth]{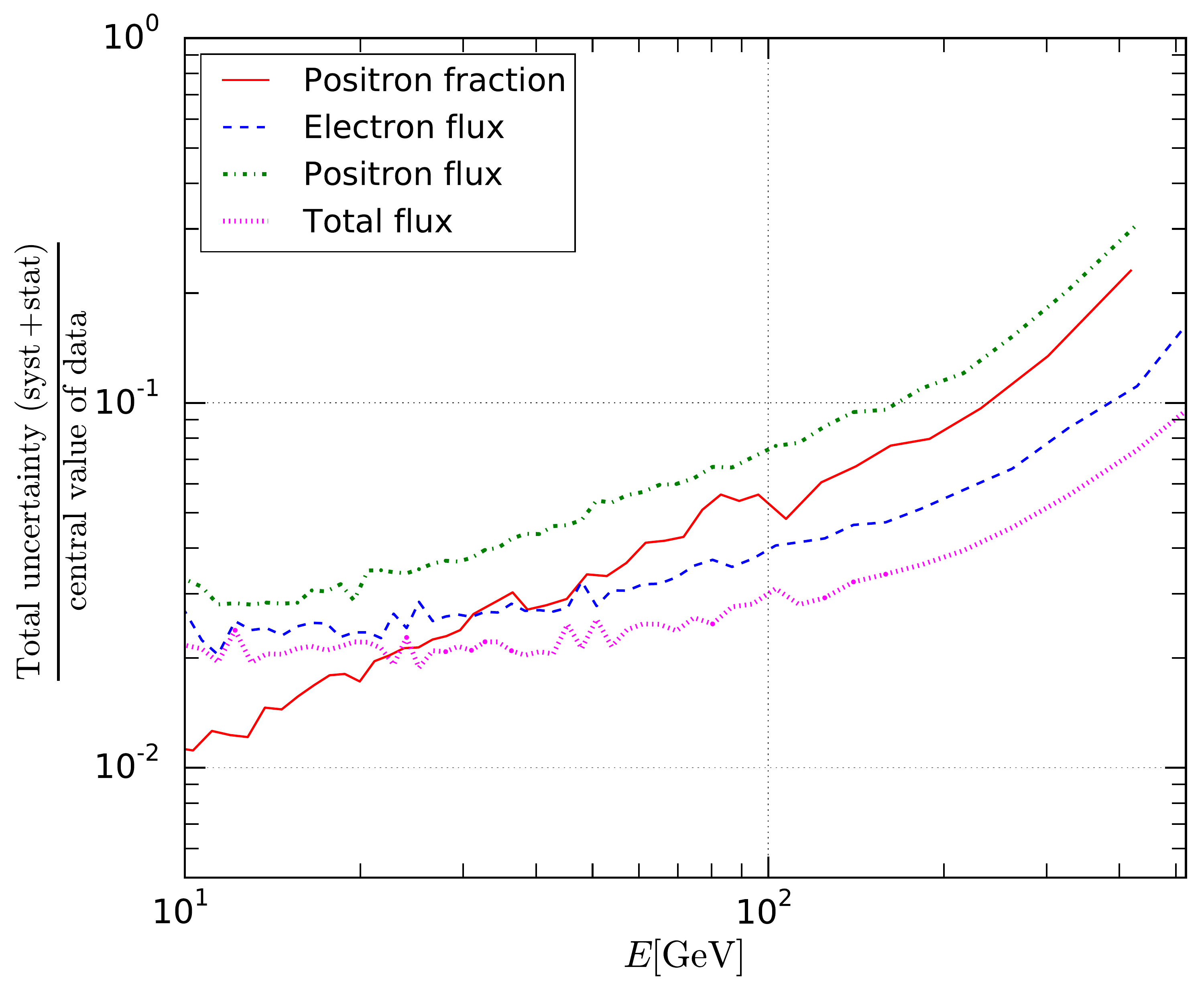}
\includegraphics[width=0.49\textwidth]{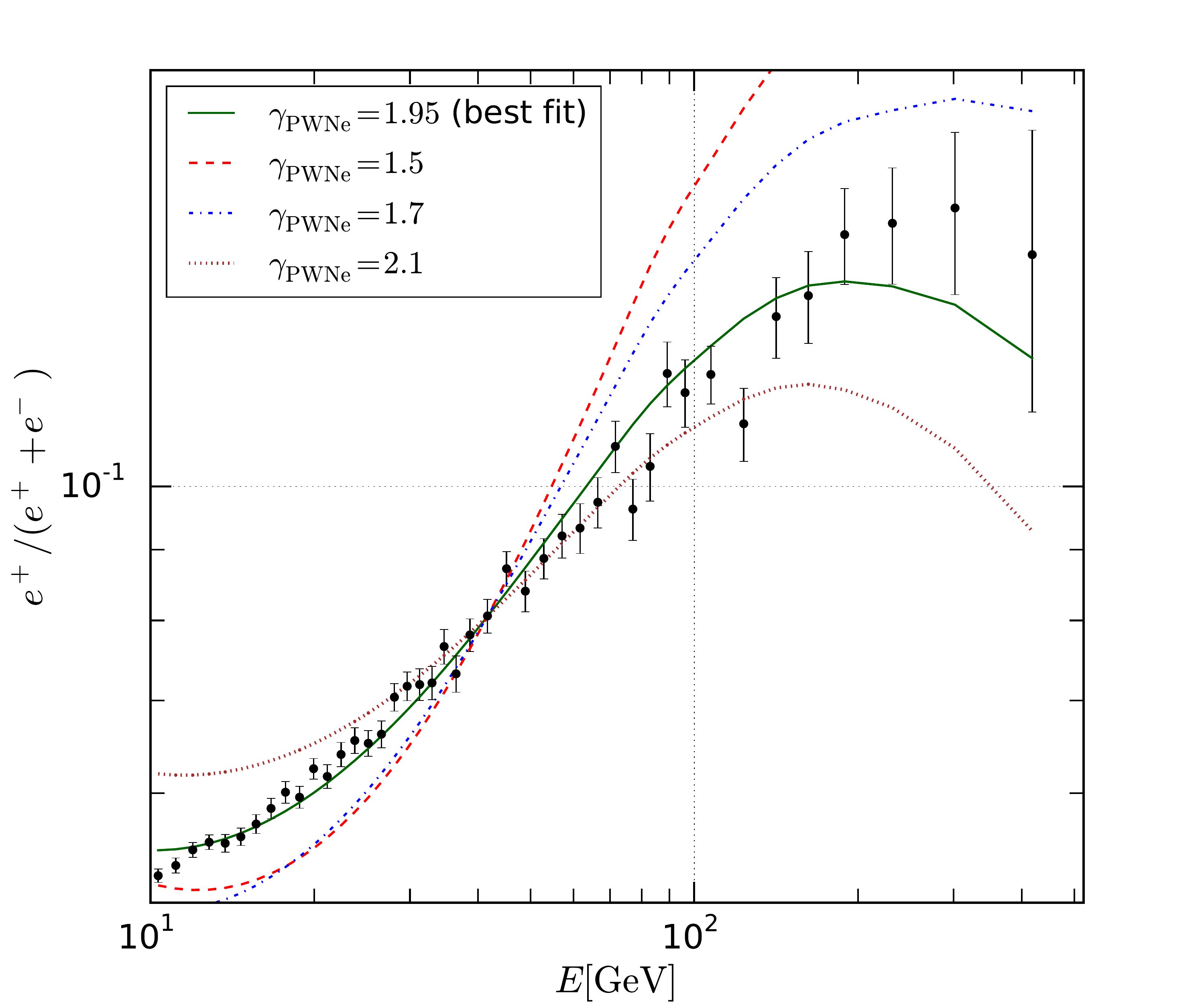}
\includegraphics[width=0.49\textwidth]{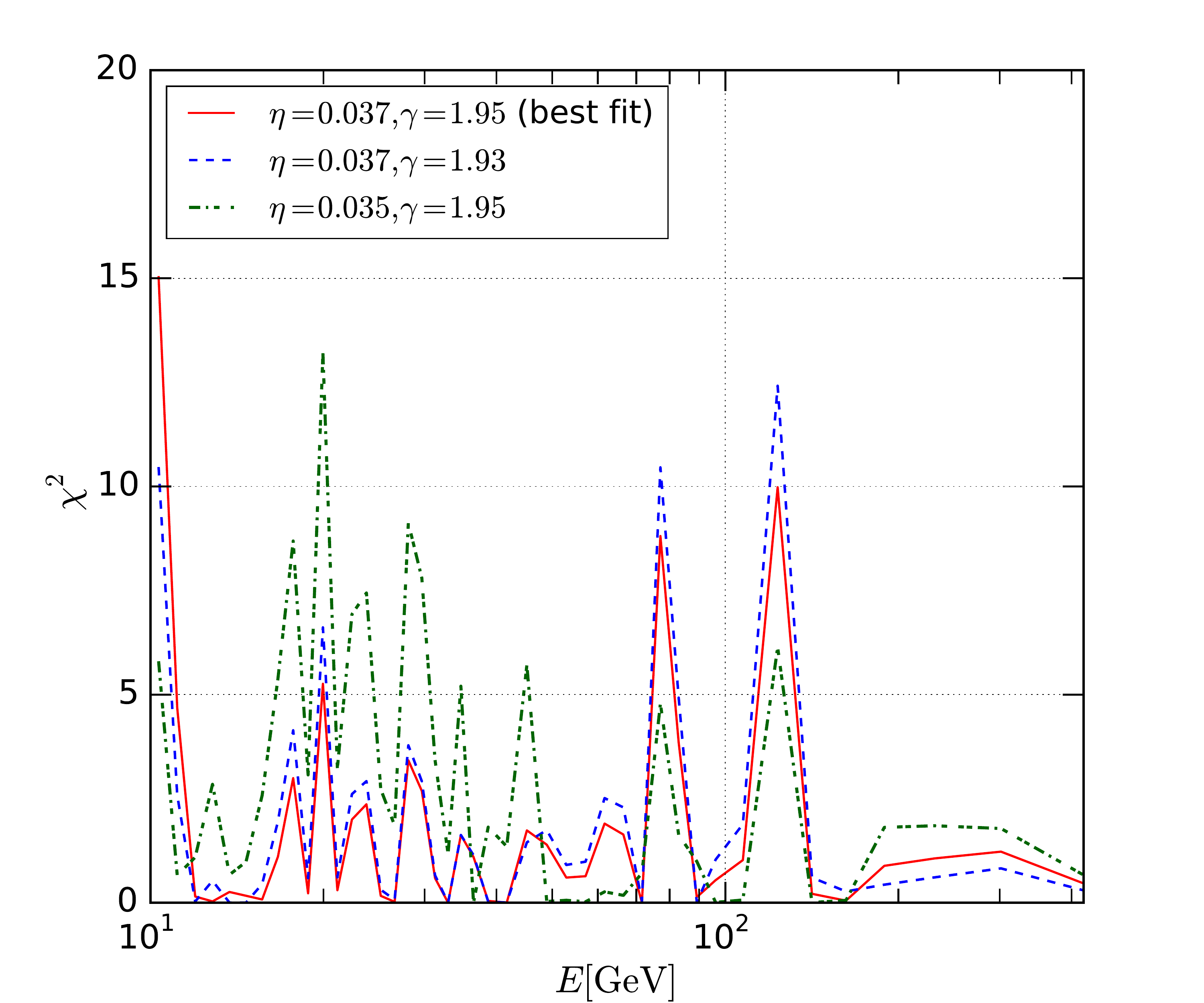}
\caption{ {\it Top left panel}: the total (statistical plus systematics) uncertainties divided by the central value of the data for the four observables measured by AMS-02, are shown as functions of the energy. {\it Top right panel}: the AMS-02 positron fraction data are reported together with predictions of the {\it astro model} for different values of the $\gamma_{\mathrm{PWNe}}$ parameter, as reported in the boxed inset. The curves are normalized in order to be equal at 41.66 GeV. {\it Bottom panel}: the $\chi^2$ associated with the positron fraction (obtained in the fit of the positron fraction and total flux datasets) is shown, as a function of the energy, for different values of the $\eta_{\mathrm{PWNe}}$ and $\gamma_{\mathrm{PWNe}}$ parameters.}
\label{fig:astro_chi2}
\end{figure}

This fact is further investigated in Fig. \ref{fig:astro_chi2}: in the top left panel we plot the relative uncertainty associated with the different datasets as a function of the energy. The curve for the positron fraction is steeper than the other curves and, for energies below 30 GeV  it is up to a factor of 3 lower than the uncertainties characterizing the other datasets. Notice also that the relative uncertainty for the positron fraction is always smaller than the one for the positron flux, in the whole energy range. Since positrons are the more relevant observables to study dark matter, and since the positron fraction is experimentally better determined, we decide to adopt this as the key observables for our analysis. And since this observables appears to have some tension with our pure  {\it astro} model, we investigate how the agreement can be improved. 

In this regard, it is useful to recall that, within the assumptions that characterize the {\it astro} model, the behaviour of the positron fraction is determined by the two parameters related to the emission from PWNe. While the effect of $\eta_{ \mathrm{PWNe}}$ is basically a normalization, a change in the  $\gamma_{ \mathrm{PWNe}}$ parameter reflects into a change in the slope, as shown in the top right panel of Fig. \ref{fig:astro_chi2}. It is quite evident that deviations of $\gamma_{ \mathrm{PWNe}}$ from the best-fit value 1.95 are strongly constrained by the large number of data points between 20 and 100 GeV. And obviously, once that $\gamma_{ \mathrm{PWNe}}$ is set, fluctuations in the normalization $\eta_{ \mathrm{PWNe}}$ away from the best fit value 0.037 are not possible.     

To illustrate how small variations in these two parameters affect the quality of the fit, we plot in the bottom panel of Fig. \ref{fig:astro_chi2}, for each data point of the positron fraction, the chi-square that is associated with the best-fit configuration of the {\it astro} model (for the combined fit of the positron fraction and of the total flux) together with the chi-square for slightly different values of the parameters of the PWNe. As it can be seen, the data points giving the maximal contribution to the total chi-square are the two at the lowest energies and the two at 77.1 GeV and 123.31 GeV (this arises from the wiggly features that the positron fraction possesses at these energies). It is evident that, in the  {\it astro} model, every change of parameters that goes in the direction of lowering the chi-square at low energies has the effect of augmenting the chi-square at high energies. This brings us to the conclusion that the modeling is not sufficient to predict a positron fraction which can be in agreement with AMS data at low and high energies at the same time. 

Starting from this observation, in the next Sections, we will extend our model to investigate whether, and under which conditions, adding an additional positron component to the {\it astro} model can improve the interpretation of AMS-02 data. Our investigation will be oriented in two directions: a contribution coming from dark matter annihilation; a less exotic astrophysical interpretations in terms of additional pulsars.

Before moving on to the discussion, we remark that all the results that are shown in the following Sections are obtained by fitting only the positron fraction and the total $e^+ + e^-$ flux. The reason for this choice is that, as it has been discussed in the previous paragraphs, the issue in reproducing AMS-02 regards solely the positron fraction. As we have commented in \cite{DiMauro:2014iia}  the four observables are the result of independent experimental analyses and therefore could be considered to be, at least to some extent, independent. Nevertheless, since under a theoretical point of view only two of them are {\it really} independent, by fitting all the four datasets, a certain degree of redundancy enters in the analysis and could hide the information concerning the positron fraction. This can be clearly seen in the results that we report in Table \ref{tab:astro}, where the fit with four datasets is associated to a low chi-square despite the tension of the model with the measured positron fraction. \\
The impact that the choice of the datasets included in the fit has on the results of some of the analyses that will be discussed in the following is discussed in Appendix \ref{app:datasets}.

\subsection{Adding dark matter to the picture}

%The previous analysis for the {\it astro model} represents our baseline modeling, over which we now  build the study of the information that can be derived on a DM contribution from the AMS-02 data. 
We fit  the positron fraction and the total flux within a model that consists of all the primary and secondary astrophysical contributions discussed in the previous paragraph {\it plus} $e^{\pm}$ fluxes produced by DM pair annihilation or decay. For definiteness, we refer to this modeling as the {\it astro+DM model}. 
The free parameters are now 8: the same $Q_{0,{\rm SNRs}}$, $\gamma_{\rm SNRs}$, $N_{\rm Vela}$, $\eta_{\rm PWNe}$, $\gamma_{\rm PWNe}$, $\phi$ already introduced before, plus the two parameters characterizing the DM particle, namely its mass $m_{DM}$ and annihilation cross-section $\sigmav$ (or decay rate $\Gamma_{\rm dec}=1/\tau$). 

Our investigation is twofold, being aimed at: {\it i)} deriving bounds on the DM annihilation/decay rate as a function of the DM mass (we will perform this analysis in a raster scan of the DM mass); {\it ii)} determining whether AMS-02 data require/prefer/allow a DM contribution.
In both cases, our underlying assumption is that the DM contribution to the electron and positron fluxes occurs in the background of emission from astrophysical sources (SNRs and PWNe): this is likely the more plausible situation, and we wish to investigate whether, to what extent and how robustly a DM leptonic emission can be present in the CR fluxes. Our analysis will not restrict the DM contribution to be neither the dominant one, or the subdominant one a priori: we will just leave free the contribution from all sources (DM, SNRs and PWNe) and adapt the modeling to the data with the same MCMC procedure discussed above.

\subsubsection{Constraints on dark matter annihilation/decay }

\begin{figure}
\centering
\includegraphics[width=0.49 \textwidth]{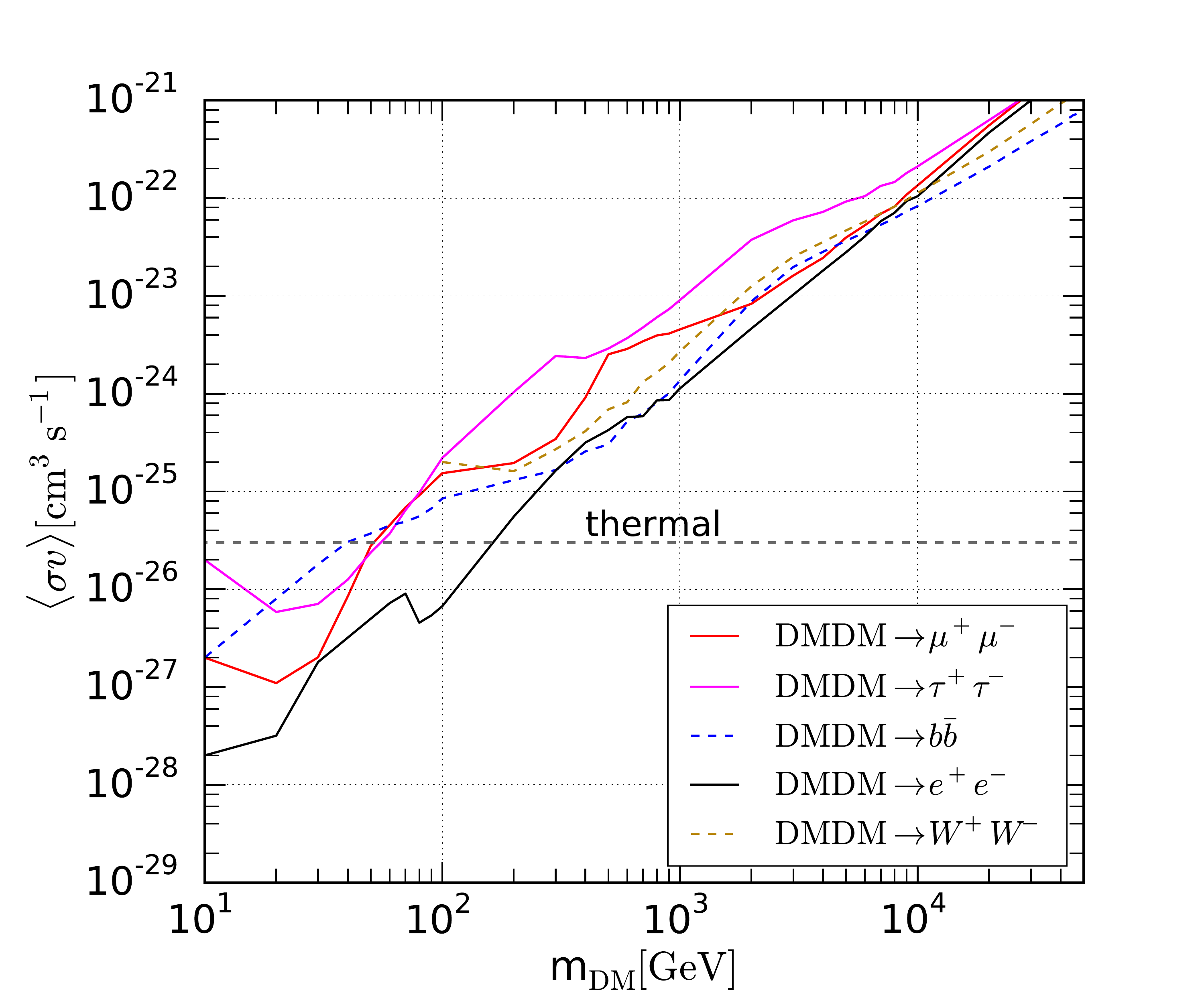}
\includegraphics[width=0.49 \textwidth]{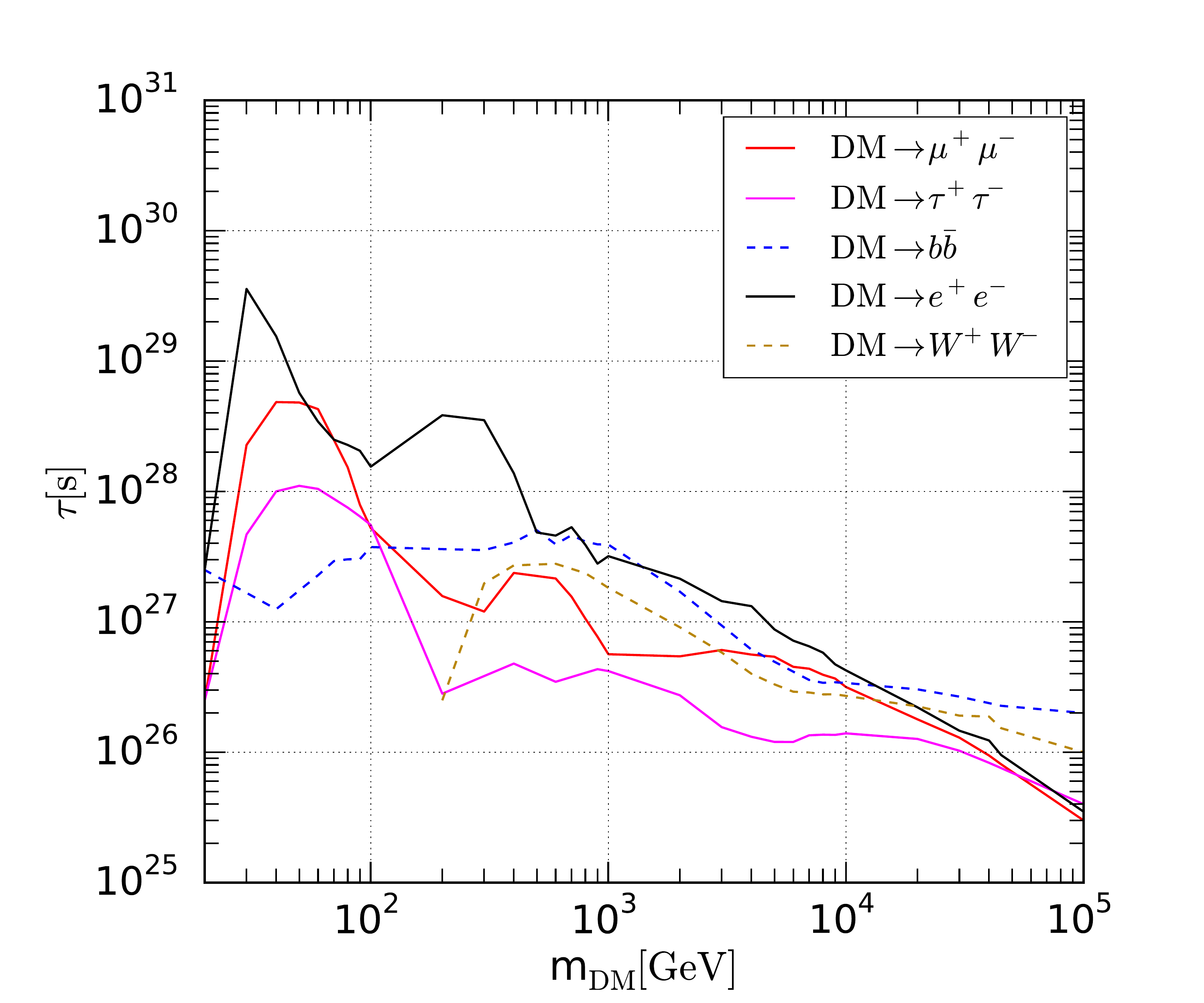}
\caption{Constraints on the DM annihilation cross section $\langle \sigma v \rangle$ (left panel) and DM lifetime $\tau$ (right panel), for various production channels (as reported in the insets) in the {\it astro+DM model}.
The constraints are computed at the $2\sigma$ C.L., and refer to the {\sc Med} propagation model and Einasto DM density profile in our Galaxy.}
\label{fig:UL}
\end{figure} 

Let us move to the determination of upper bounds on the DM annihilation cross section $\sigmav$ (or lower bounds on the decay lifetime $\tau$) within a realistic full model for the astrophysical emission of primary and secondary $e^{\pm}$. We proceed as follows: for every DM annihilation/decay channel and for each value of the DM mass, we perform a MCMC sampling of the space defined by the 7 free parameters that characterize our model (6 astrophysical parameters plus $\sigmav$ or $\tau$). By marginalizing over the astrophysical-sources parameters, we derive the posterior distribution functions for the DM annihilation cross section (or lifetime, in case of decaying DM), from which  $2\sigma$ C.L. upper limits are obtained. This refers to the values $\langle \sigma v \rangle_{\mathrm{lim}}$ and $\tau_{\mathrm{lim}}$ for which $P(\langle \sigma v \rangle \le \langle \sigma v \rangle_{\mathrm{lim}}) = 97.73\%$ and $P(\tau \ge \tau_{\mathrm{lim}}) = 97.73\%$, being P the posterior distribution function of the DM annihilation cross section or lifetime. The results for both annihilating and decaying dark matter are shown in Fig.~\ref{fig:UL}.  

The bounds are quite constraining, especially in the low DM-mass region. For the annihilating case,
the softer hadronic channels (here represented by the $b\bar b$ channel) exclude thermal annihilation cross section for DM masses below 40 GeV. The same occurs for the harder $\mu^+\mu^-$ and $\tau^+\tau^-$ channels, while for $e^+e^-$ channel, which has the most prominent spectral feature 
 \cite{Delahaye:2007fr}, the bound on thermal cross sections increases at 200 GeV. 

For the leptonic annihilation channels, we notice two features. The first is that for lighter DM, the bounds can be quite strong. Annihilation into leptons, especially the direct annihilation into a $e^+e^-$ and to some extent also $\mu^+\mu^-$, have a hard energy spectrum with a characteristic feature \cite{Delahaye:2007fr}, with a sharp fall-off: when the DM mass is below about 100 GeV, the feature occurs in the lower-energy portion of the positron flux, where the astrophysical contribution is dominated by the soft and smooth secondary flux which, as already mentioned, reproduces remarkably well the AMS-02 data. In this case, significantly strong bounds are obtained. The limits obtained here are on average in good agreement with the ones that have been derived in Refs.~\cite{Ibarra:2013zia,Bergstrom:2013jra}, even though the astrophysical modeling in our analysis and in Refs.~\cite{Ibarra:2013zia,Bergstrom:2013jra} differ.

The second feature observable in the bounds of Fig.~\ref{fig:UL} for the $e^+e^-$ and $\mu^+\mu^-$ channels is represented by a series of bumps that appear in the constraints. This corresponds to the fact that for some DM mass values, a contribution from DM annihilation or decay 
is actually preferred by the fit, and therefore the bounds weaken. These facts are investigated in the next Section.  

\subsubsection{Allowed regions in the DM parameters space}
\begin{table}[t]
\centering
\footnotesize
\begin{tabular}{|c|c|c|c|c|c|}
\hline
\multicolumn{6}{|c|}{{\bf Annihilating DM}\rule{0pt}{20pt}} \\
\hline
Parameter & $\bf e^+ e^-$ & $\mu^+\mu^-$ & $\tau^+ \tau^-$ & $b\bar{b}$ & $W^+W^-$ \\
\hline
$\eta_{PWNe}$ & $0.032\substack{+0.002 \\ -0.002} $& $0.028\substack{+0.002 \\ -0.005}$ & $0.011\substack{+0.005 \\ -0.001}$ & $0.006\substack{+0.015 \\ -0.001}$ & $0.006\substack{+0.011 \\ -0.001}$ \\ 
\hline
$\gamma_{PWNe}$ & $1.87\substack{+0.05 \\ -0.05} $& $1.76\substack{+0.09 \\ -0.20} $& $1.23\substack{+0.33 \\ -0.23}$ & $1.77\substack{+0.19 \\ -0.69}$ & $1.72\substack{+0.27 \\ -0.68} $ \\
\hline
$Q_{0, SNRs} [10^{50}$ erg/s] & $1.13\substack{+0.12 \\ -0.09} $&$1.24\substack{+0.10 \\ -0.18}$ &$1.16\substack{+0.14 \\ -0.05}$ & $1.40\substack{+0.11 \\ -0.14}$ & $1.33\substack{+0.12 \\ -0.11}$ \\
\hline
$\gamma_{SNRs}$ &$ 2.22\substack{+0.02 \\ -0.01}$  & $2.24\substack{+0.01 \\ -0.03}$ & $2.23\substack{+0.02\\ -0.01}$ & $2.26\substack{+0.01 \\ -0.02}$ & $2.25\substack{+0.02 \\ -0.01}$ \\
\hline
$N_{Vela}$ & $0.80\substack{+0.19 \\ -0.17}$ & $0.74\substack{+0.24 \\ -0.20}$ & $0.88\substack{+0.14 \\ -0.20}$& $0.84\substack{+0.22 \\ -0.15}$&$0.81\substack{+0.22 \\ -0.17}$\\
\hline
$m_{DM}$ [GeV] & $50\substack{+1 \\ -4} $& $88\substack{+31 \\ -9}$ & $635\substack{+73\\ -195}$ &$ 39572\substack{+10351\\ -28792} $& $24759\substack{+22964 \\ -14907}$ \\   
\hline
$\langle \sigma v \rangle $ [cm$^3$s$^{-1}$]&$ 5.6\substack{+2.2 \\ -2.6} \times 10^{-27} $&  $7.9\substack{+12.6 \\ -3.4} \times 10^{-26} $& $7.2\substack{+1.4 \\ -3.5}\times 10^{-24}$ &$ 9.5\substack{+0.5 \\ -8.4}\times 10^{-22} $&$ 7.2\substack{+11.5 \\ -5.7}\times 10^{-22}$ \\
\hline
$\chi^2/85~{\mathrm {d.o.f.}}$ & 1.13 & 0.98 & 1.05 & 1.24 & 1.18 \\
\hline
%$p$-value(sign.) & $0.20(1.3\sigma)$ & $1.1\times10^{-2}(2.5\sigma)$ & $5.8\times10^{-2}(1.9\sigma)$ & $0.33(1.0\sigma)$ & $0.31(1.0\sigma)$ \\
%\hline
\multicolumn{6}{|c|}{{\bf Decaying DM}\rule{0pt}{20pt}} \\
\hline
Parameter & $\bf e^+ e^-$ & $\mu^+\mu^-$ & $\tau^+ \tau^-$ & $b\bar{b}$ & $W^+W^-$ \\
\hline
$\eta_{PWNe}$ & $0.032\substack{+0.002 \\ -0.002} $& $0.026\substack{+0.002 \\ -0.004}$ & $0.011\substack{+0.005 \\ -0.001}$ & $0.006\substack{+0.004 \\ -0.001}$ & $0.006\substack{+0.013 \\ -0.001}$ \\ 
\hline
$\gamma_{PWNe}$ & $1.88\substack{+0.04 \\ -0.05} $& $1.71\substack{+0.10 \\ -0.17} $& $1.15\substack{+0.33 \\ -0.23}$ & $1.82\substack{+0.08 \\ -0.21}$ & $1.66\substack{+0.25 \\ -0.66} $ \\
\hline
$Q_{0, SNRs} [10^{50}$ erg/s] & $1.16\substack{+0.10 \\ -0.11} $&$1.20\substack{+0.10 \\ -0.10}$ &$1.15\substack{+0.43 \\ -0.15}$ & $1.44\substack{+0.14 \\ -0.10}$ & $1.31\substack{+0.13 \\ -0.08}$ \\
\hline
$\gamma_{SNRs}$ &$ 2.23\substack{+0.01 \\ -0.02}$  & $2.23\substack{+0.02 \\ -0.02}$ & $2.23\substack{+0.01\\ -0.01}$ & $2.26\substack{+0.03 \\ -0.01}$ & $2.25\substack{+0.02 \\ -0.01}$ \\
\hline
$N_{Vela}$ & $0.85\substack{+0.14 \\ -0.21}$ & $0.73\substack{+0.19 \\ -0.20}$ & $0.90\substack{+0.18 \\ -0.22}$& $0.81\substack{+0.20 \\ -0.13}$&$0.79\substack{+0.23 \\ -0.13}$\\
\hline
$m_{DM}$ [GeV] & $102\substack{+2 \\ -9} $& $189\substack{+63 \\ -23}$ & $1211\substack{+368\\ -475}$ &$ 56857\substack{+29977\\ -28818} $& $45448\substack{+2414 \\ -26575}$ \\   
\hline
$\tau $ [s]&$ 2.4\substack{+1.6 \\ -0.7} \times 10^{28} $&  $2.2\substack{+1.2 \\ -0.8} \times 10^{27} $& $2.2\substack{+0.1 \\ -0.2}\times 10^{26}$ & $1.1\substack{+0.4 \\ -0.2}\times 10^{26} $& $0.9\substack{+1.1 \\ -0.08}\times 10^{26} $\\
\hline
$\chi^2/85~{\rm d.o.f.}$ & 1.14 & 1.00 & 1.04 & 1.28 & 1.18 \\
\hline
%$p$-value(sign.) & $0.21(1.2\sigma)$ & $2.3\times10^{-2}(2.2\sigma)$ & $5.2\times10^{-2}(1.9\sigma)$ & $0.33(1.0\sigma)$ & $0.24(1.2\sigma)$ \\
%\hline
\end{tabular}
\caption{Best-fit configurations for the {\it astro+DM model}.}
\label{tab:bestfits}
\end{table}

\begin{figure}
\centering
\includegraphics[scale=0.45]{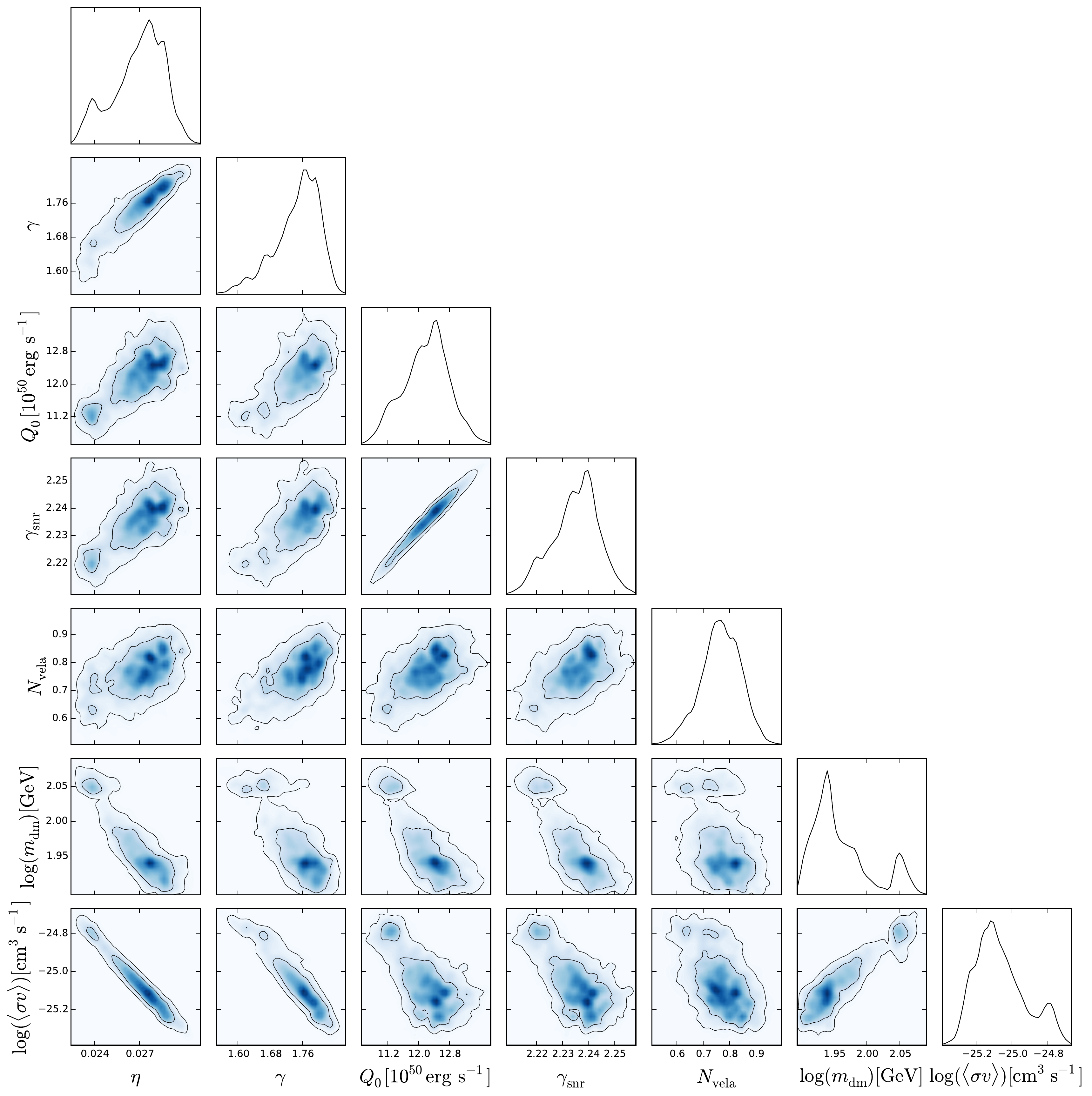}
\caption{Triangular plot for the fit of the parameters of the {\it astro+DM model} to AMS-02 data, for a DM annihilating into the $\mu^+\mu^-$ channel. The contours refer to $1\sigma$, $2\sigma$ and $3\sigma$ C.L. allowed regions. The plots along the diagonal show the posterior distribution for each parameter.}
\label{fig:tri_mu}
\end{figure}
\begin{figure}
\centering
\includegraphics[width=0.49 \textwidth]{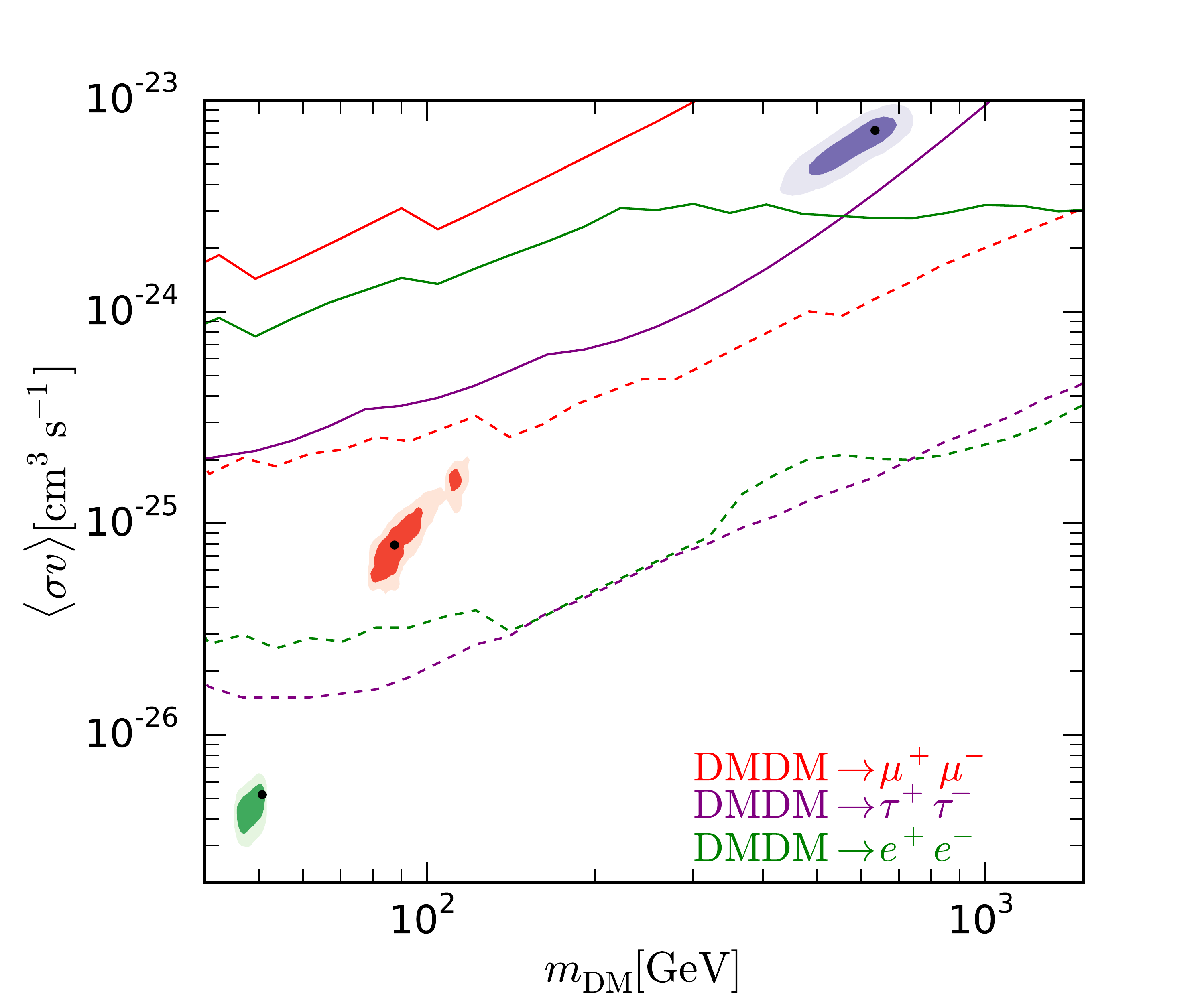}
\includegraphics[width=0.49 \textwidth]{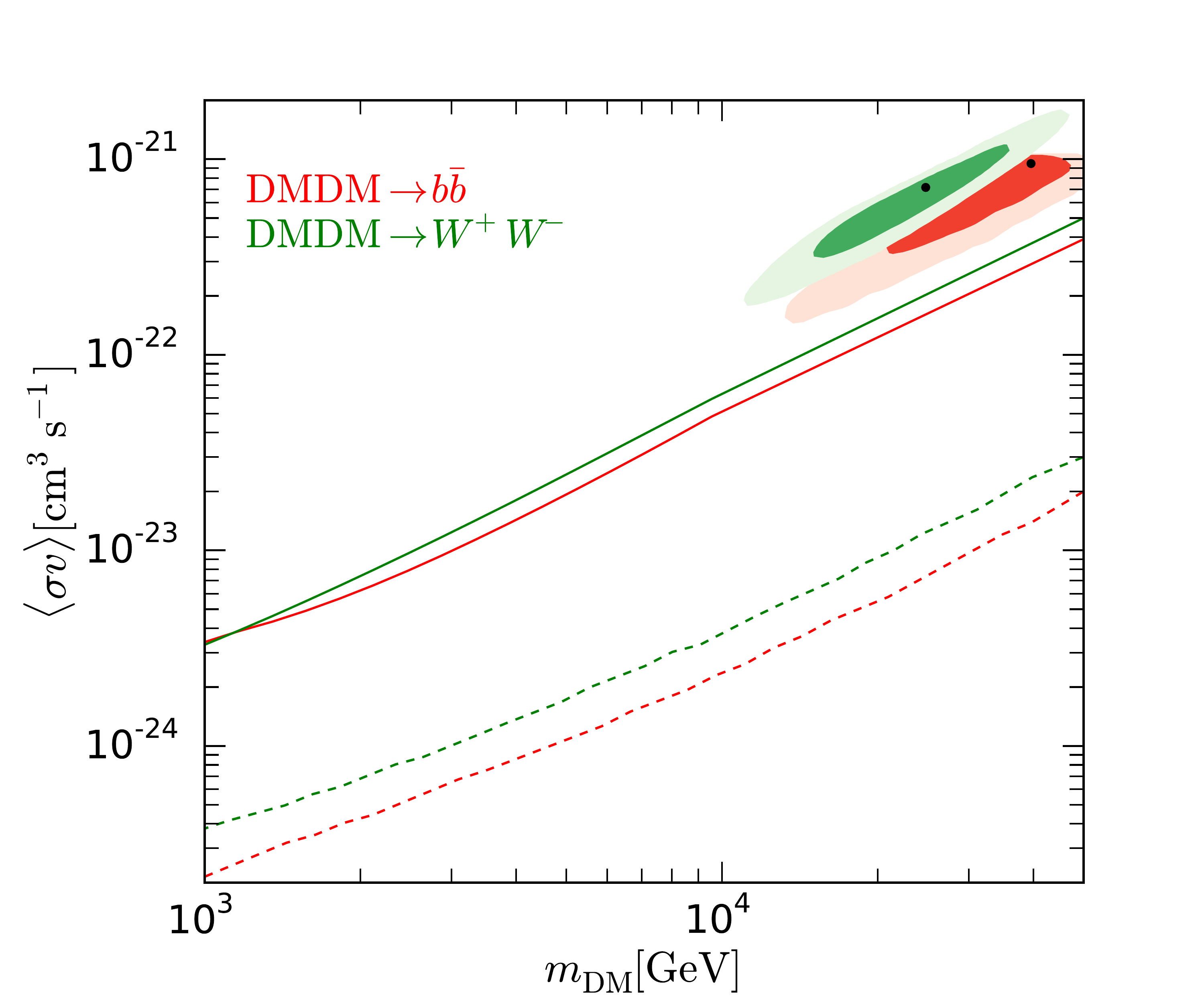}
\includegraphics[width=0.49 \textwidth]{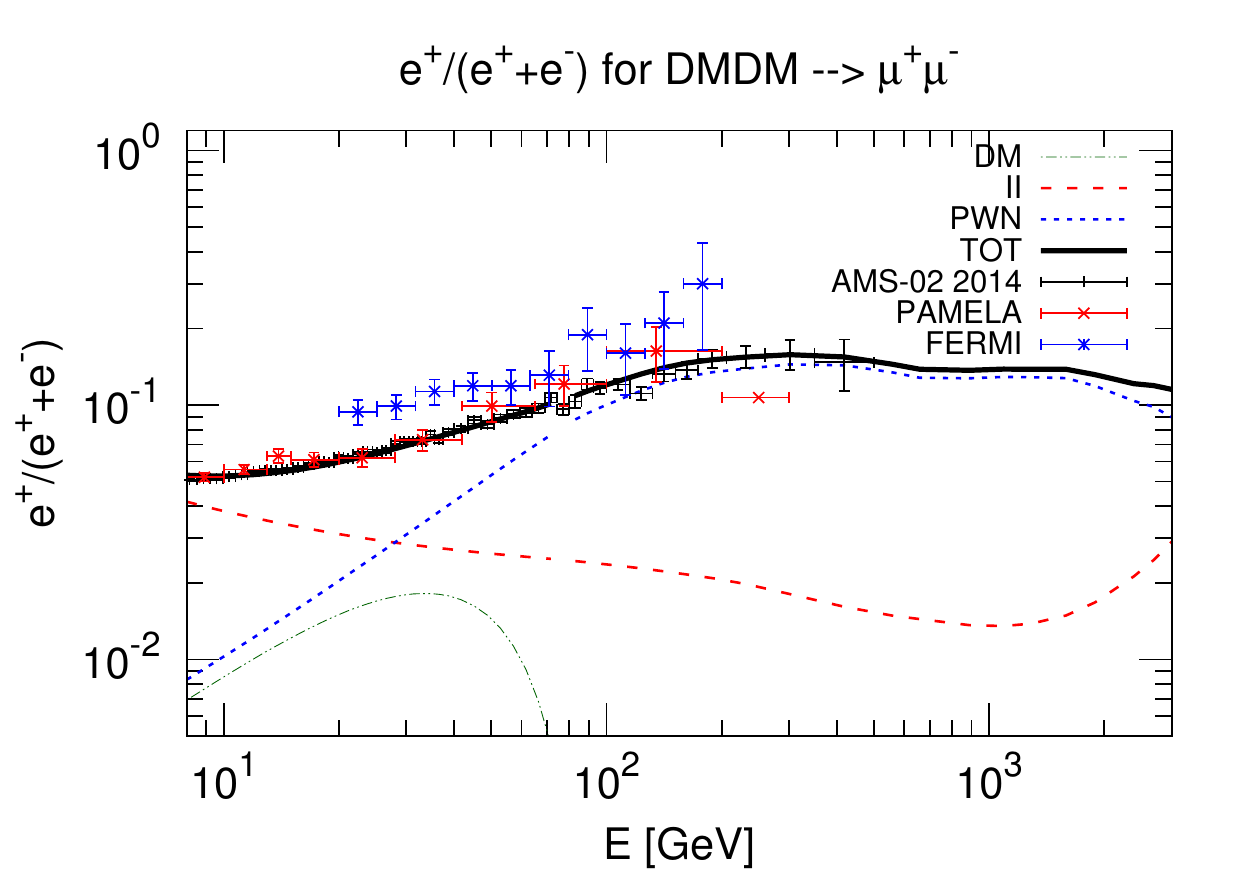}
\includegraphics[width=0.49 \textwidth]{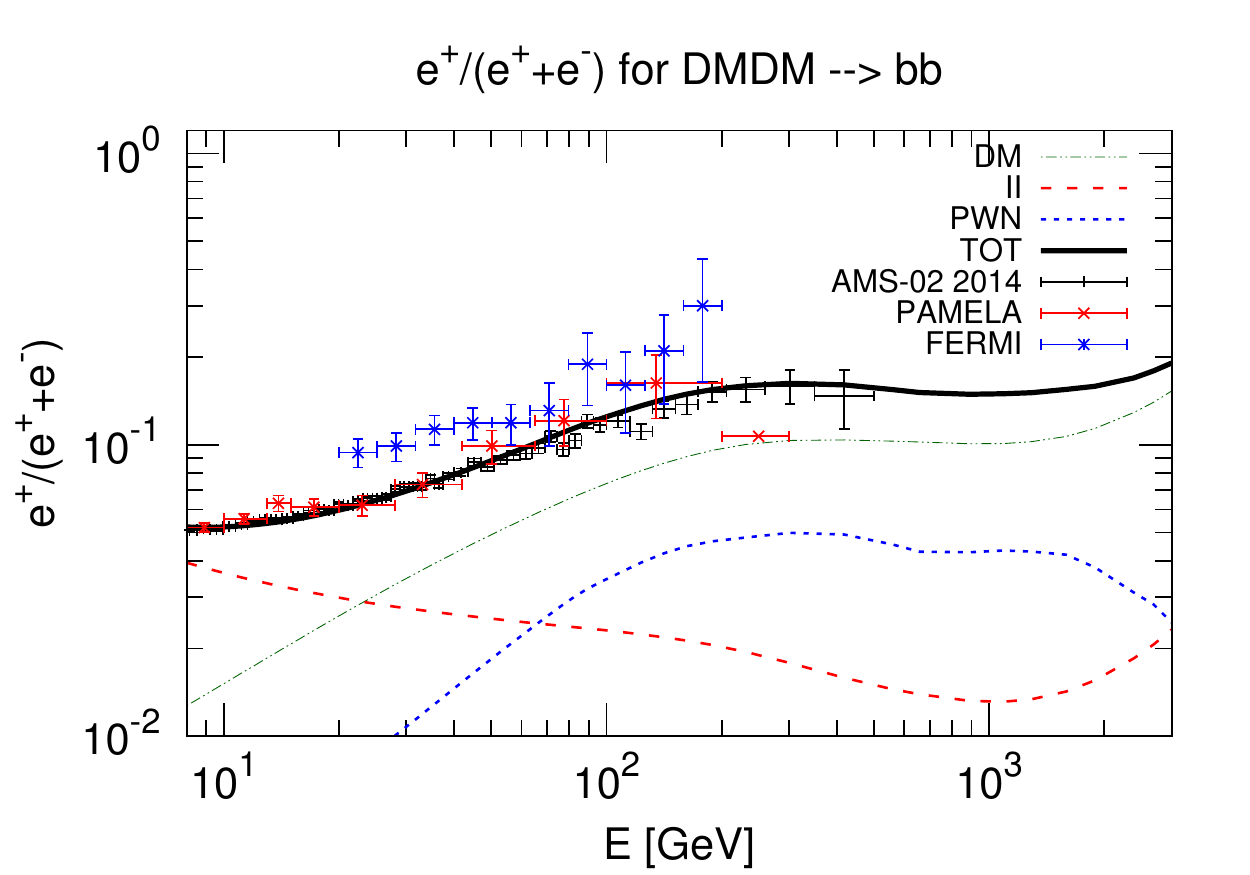}
\caption{Upper row: allowed regions in the $(m_{DM},\sigmav)$
for a DM particle annihilating into leptonic ({\it left panel}) or hadronic ({\it right panel}) channels. For every channel, the different shadings correspond to the $1\sigma$ and $2\sigma$ contour regions, while the solid and dashed lines represent, respectively, the conservative and optimistic gamma-ray upper limits derived in Refs.~\cite{Calore:2013yia,DiMauro:2015tfa}. In the lower row, the contributions to the positron fraction  associated to the best-fit configurations for a  DM annihilating in $\mu^+ \mu^-$ ({\it left panel}) and $b \bar{b}$ ({\it right panel}) are shown.}
\label{fig:CP}
\end{figure} 

We therefore explicitly investigate  which are the configurations in the DM parameters space that can provide the best fits to AMS-02 data. Instead of a raster scan on the DM mass, we investigate the full 8-fold parameter space of the {\it astro + DM model} in order to derive the best fit configurations together with their allowed regions. The results are reported in Table \ref{tab:bestfits}, while 
the triangular plot of Fig.~\ref{fig:tri_mu} illustrates the posterior probability distributions of the parameters for the case in which DM annihilates into $\mu^+ \mu^-$. The $1 \sigma$ and $2 \sigma$ contour plots in the $(m_{DM},\langle \sigma v \rangle)$ plane for the annihilating DM case are reported in the upper row of Fig.~\ref{fig:CP} together with the bounds that have been found in Refs.~\cite{Calore:2013yia,DiMauro:2015tfa} as a result of an investigation of the different components contributing to the isotropic gamma-ray background (IGRB) measured by the {\it Fermi}-LAT experiment. Lastly, in the lower row of Fig.~\ref{fig:CP} the different contribution to the positron fraction measured by AMS-02 are shown for the best-fit configurations of the two cases of DM particles annihilating into the $\mu^+ \mu^-$ and $b\bar{b}$ channels. 

As it can be clearly seen, the fitting procedure can reach a very high resolution in determining the mass and the annihilation cross section/lifetime for the leptonic annihilation/decay channels. This is a consequence of the fact that these channels are expected to produce sharply peaked positron spectra which basically contribute to replenish the gap at intermediate energies ({\it i.e.} between 20 and 40 GeV), where the secondary flux begins to become negligible and the contribution from PWNe starts getting relevant. In this relatively small energy window, small variations in DM parameters can lead to large fluctuations of the chi-square and this makes the contour regions in the $(m_{DM},\langle \sigma v \rangle)$ plane quite small. A different situation occurs for hadronic channels, for which the DM contribution extends over several energy bins, from intermediate energies up to the last AMS-02 bins: in a large fraction of this energy range, the contribution from DM is degenerate with the contribution from PWNe and thus quite large variations in DM properties do not always translate into large fluctuations on the chi-square.  

Concerning the comparison with gamma-ray constraints on the same annihilation channels, the upper row of Fig.~\ref{fig:CP} reports the upper limits on the DM thermally averaged annihilation cross section $\langle \sigma v \rangle$ computed in Refs.~\cite{Calore:2013yia,DiMauro:2015tfa} by means of a detail modelling of the various contribution to the IGRB energy spectrum that can be derived from the measurements performed by the {\it Fermi}-LAT experiment. In particular, solid lines represent {\it conservative} upper limits as reported in Ref.~\cite{Calore:2013yia} with the $\gamma$-ray emission from extragalactic sources set to the minimal level allowed in their modeling, while dashed lines refer to {\it optimistic} upper limits derived in Ref.~\cite{DiMauro:2015tfa} from a statistical fit to the {\it Fermi}-LAT IGRB data. 
As it can be seen, only the annihilation in the $e^+e^-$ and $\mu^+\mu^-$ channels is fully consistent with the gamma-ray bounds, while for all the other channels a tension is present, even in a conservative setup.  

%The last result worth emphasizing is that DM improves the agreement in the positron fraction channel, when compared to the pure {\it astro model}. This occurs for bot the $\mu^+\mu^-$ and $\tau^+ \tau^-$ production channels, where the chi-square per data point drops (from the value of 1.9 obtained in the {\it astro model}) to 1.14 and 1.23 respectively.  In other words, for these channels, adding a DM contribution to the $e^{\pm}$ astrophysical emission indeed improves the agreement with the AMS-02 data.
%decreasing the $p$-value of the fit from 
%0.45 (about $1\sigma$) for the pure astrophysical interpretation to $6\times 10^{-3}$ ($2.7\sigma$) for the %$\mu^+\mu^-$ channel,
%as can be seen by comparing the results in Table \ref{tab:astro} and Table \ref{tab:bestfits}.

\begin{figure}
\centering
\includegraphics[width=0.49 \textwidth]{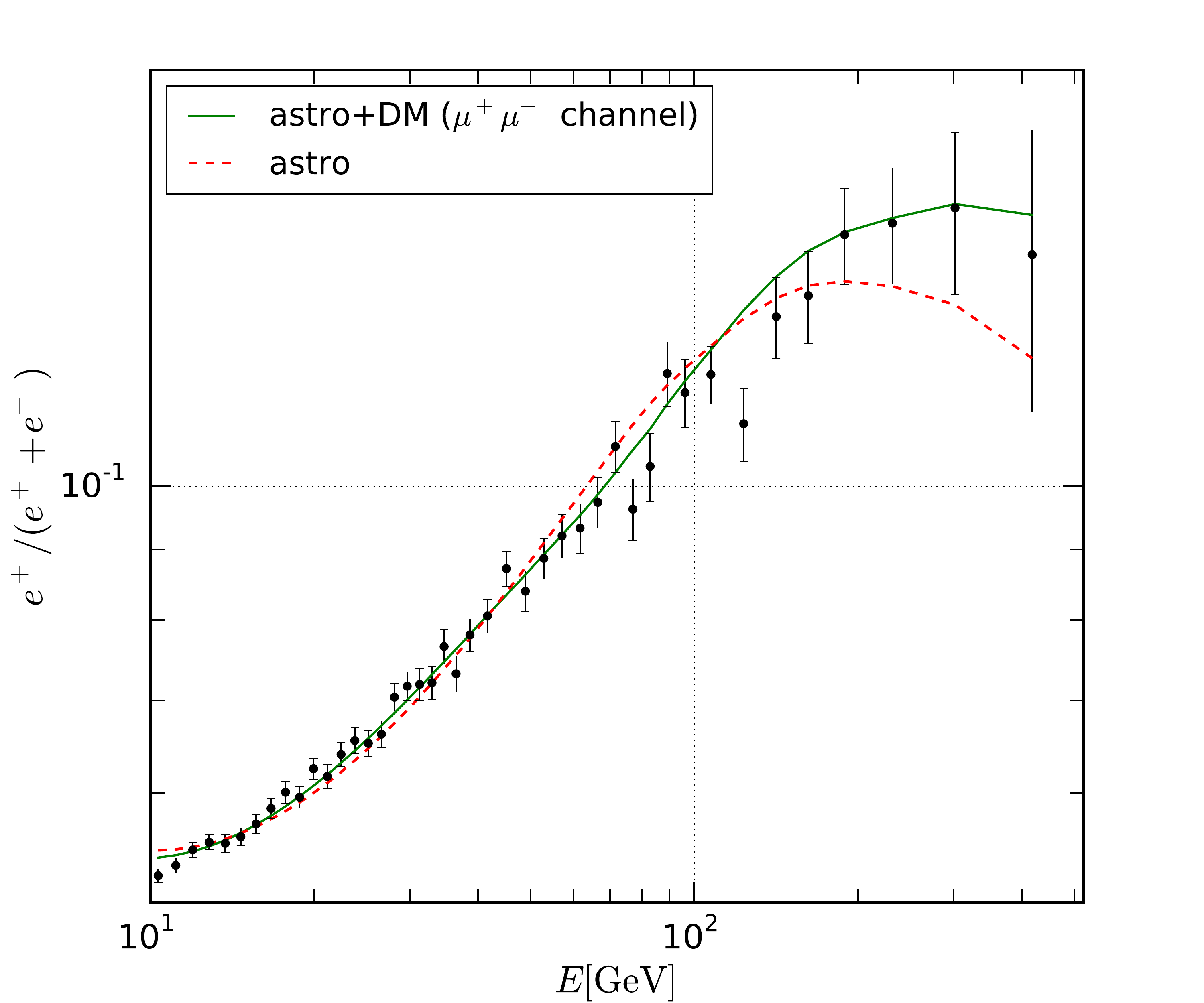}
\includegraphics[width=0.49 \textwidth]{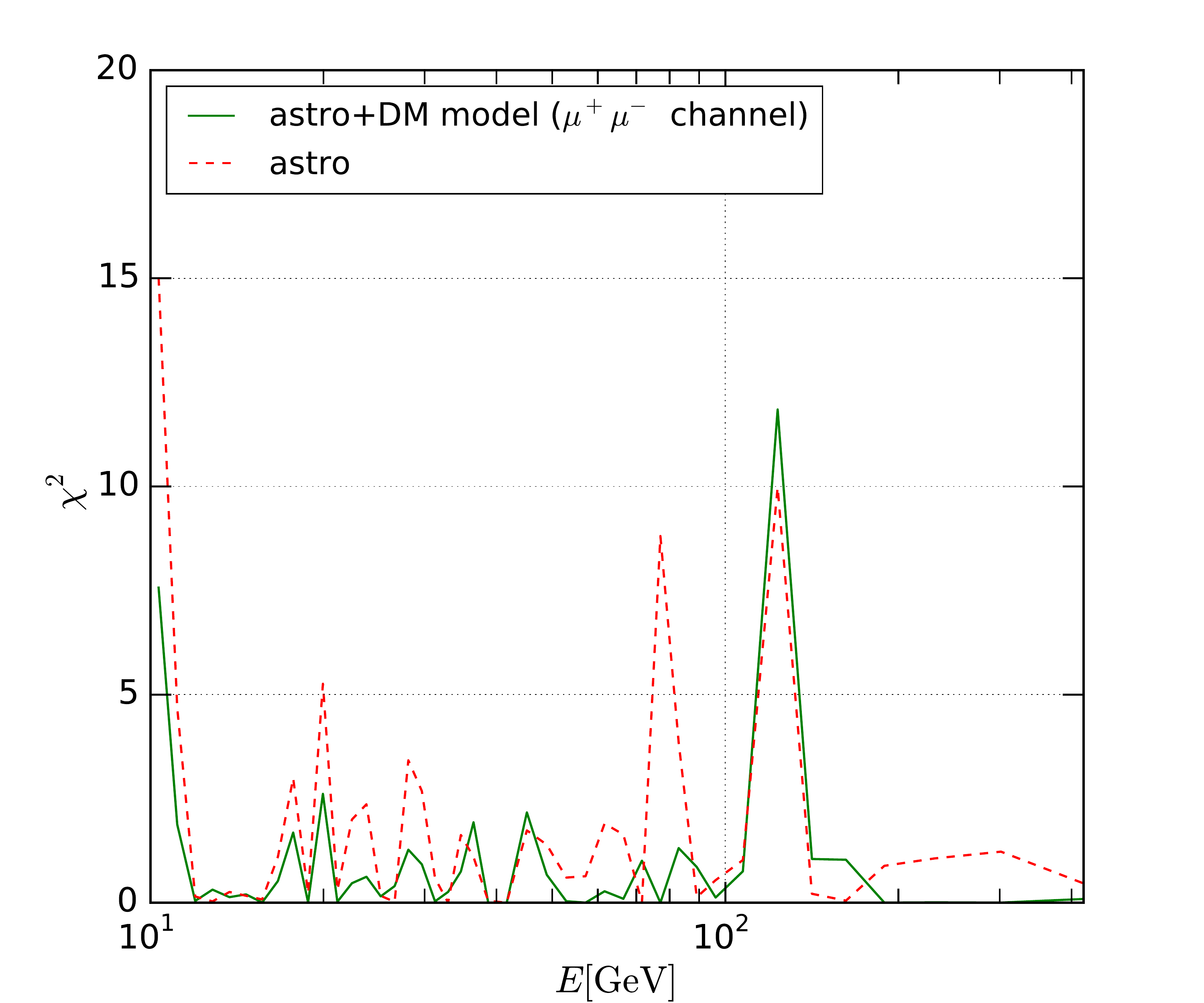}
\caption{ {\it Left panel}: positron fraction for the best-fit configurations of the {\it astro} model and of the {\it astro + DM} model ($\mu^+\mu^-$ channel), together with AMS-02 data, as a function of the energy. {\it Right panel}: the $\chi^2$ associated to the two models is shown as a function of the energy.}
\label{fig:DM_chi}
\end{figure}

It is worth to emphasize that the {\it astro + DM} model has a better agreement with AMS-02 data than the pure {\it astro} model. This is especially true for $\mu^+\mu^-$ and $\tau^+ \tau^-$ annihilation/decay channels, for which the reduced $\chi^2$ drops to values around 1 (for the {\it astro} model we had $\chi^2/{\mathrm{d.o.f}} = 1.35$). In both cases, the predicted positron fraction is in good agreement with AMS-02 data: as an example, for the annihilating DM case, the $\chi^2$ {\it per data point} of this observable is 1.00 for the $\mu^+\mu^-$ channel and 1.01 for the $\tau^+\tau^-$ channel. To give a better illustration of this agreement, we plot in Fig. \ref{fig:DM_chi} both the positron fraction for the {\it astro + DM} model compared with the one predicted by the {\it astro} model and the $\chi^2$ for each data point for the two models. As it can be clearly seen, the addition of the DM contribution brings a significant improvement of the fit that is not confined to the low-energy data points, but extends over all the energy range of AMS-02 measurements.

Finally, we notice that also for the {\it astro+DM} model, the positron fraction keeps a flat behaviour beyond the current AMS-02 data: this is a consequence of the fact that the addition of a DM component is driven to contribute to intermediate energies, and leaves the high-energy part of the positron spectrum basically unchanged from the one obtained in the pure astrophysical interpretation.

\subsubsection{Varying the astrophysical modeling}    
\label{sec:psr}

\begin{table}
\center
\begin{tabular}{|c|c|}
\hline
Parameter& central value $\pm 1 \sigma$ \\
\hline
$\eta_{PWNe}$ & $0.05\substack{+0.04 \\ -0.03}$ \\ 
\hline
$\gamma_{PWNe}$ & $ 2.3\substack{+0.1 \\ -1.0} $ \\
\hline
$Q_{0, SNRs} [10^{50}$ erg/s] & $1.10\substack{+0.02 \\ -0.06}$ \\
\hline
$\gamma_{SNRs}$ & $2.22 \substack{+0.01 \\ - 0.02}$ \\
\hline
$N_{Vela}$ & $0.8 \substack{+0.2 \\ -0.2} $\\
\hline
$d_{psr}$ [kpc] & $ 0.63 \substack {+0.26 \\ -0.24}$ \\
\hline
$T_{psr}$ [kyr] & $1110 \substack{+873 \\ -610 } $ \\
\hline
$\eta_{psr}$ & $0.47 \substack{+0.23\\-0.17} $\\
\hline
\multicolumn{2}{|c|}{$\chi^2 /84~\mathrm{d.o.f.}$ = 0.85}\\
\hline 
\end{tabular}
\caption{Best fit for the {\it astro model} with an additional PWN. For every parameter the best-fit value is reported together with the upper and lower $1\sigma$ uncertainty. The value of $\eta_{psr}$ has been derived assuming a spin-down luminosity for the additional PWN equal to $\dot{E}=10^{34}$ erg/s, for definiteness.}
\label{tab:psr}
 \end{table} 
 %%%%% revised version %%%%%%%%%%%
 %\begin{table}
%\center
%\begin{tabular}{|c|c|c|c|c|}
%\hline
%$\chi^2_{\mathrm{tot}}$ (181 d.o.f) & $\chi^2_{e^+}$ (48 pts) & $\chi^2_{e^-}$ (49 pts) & $\chi^2_{\mathrm{sum}}$ (50 pts) & $\chi^2_{\mathrm{pf}}$ (43 pts) \\
%\hline
%131.27 & 26.36 & 31.72 & 34.82 & 38.37 \\
%\hline
%\end{tabular}
%\end{table}
 %%%%%%%%%%%%%%%%%%%%%%%%

In this Section we explore the consequences of relaxing some of the assumptions that, in the previous Sections, have been made about the contribution from PWNe.  This should allow us to understand to what extent the remarkable agreement that we have found in the {\it astro+DM model} can be interpreted as a genuine hint of DM, capable of surviving even within a more complex modeling of the PWNe contribution. 

As a first analysis, we carry out a simple exercise to assess what would be the properties of a putative pulsar that, added to the ones of the ATNF catalogue, would be able to provide an agreement to AMS-02 data as good as the one obtained with the addition of DM. We are therefore trying to replace the DM electron-positron production with a generic additional PWN emission, to see if this can fully mimic a DM contributions and what are the properties required for such a pulsar.
To this aim, we perform a fit to AMS-02 data within a version of the {\it astro model}, modified in such a way to include an additional PWN, whose distance $d_{\rm psr}$, age $T_{\rm psr}$ and total energy emitted in the form of $e^{\pm}$ pairs $\eta_{\rm psr} W_{0,{\rm psr}}$ are to be determined by a fit to the data.
This brings the total number of free parameters in our fit to 9: $Q_{0,{\rm SNRs}}$, $\gamma_{\rm SNRs}$, $N_{\rm Vela}$, $\eta_{\rm PWNe}$, $\gamma_{\rm PWNe}$, $\phi$, $d_{\rm psr}$,  $T_{\rm psr}$ and $\eta_{\rm psr} W_{0,{\rm psr}}$.

\begin{figure}[t]
\center
\includegraphics[width=0.49 \textwidth]{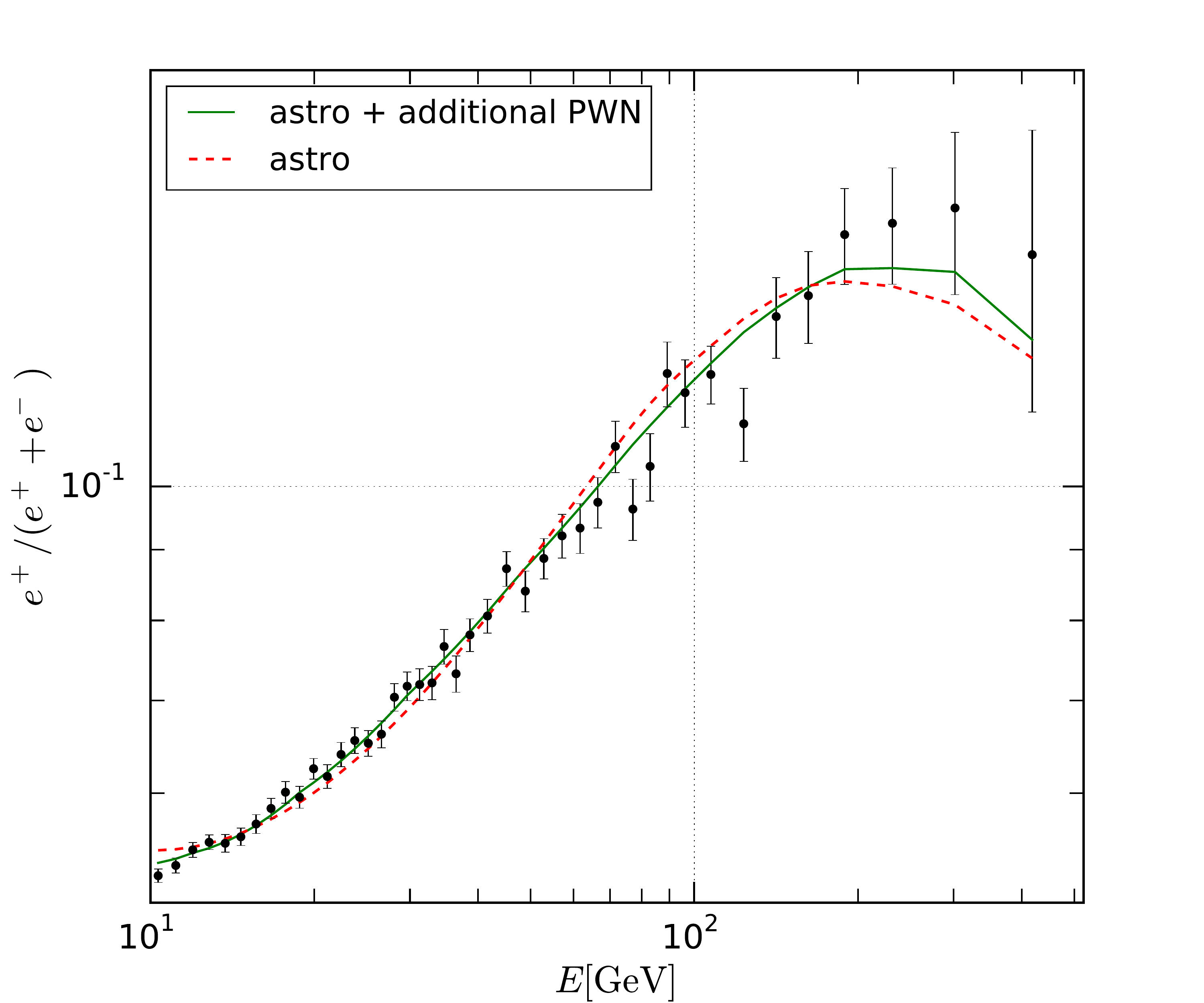}
\includegraphics[width=0.49 \textwidth]{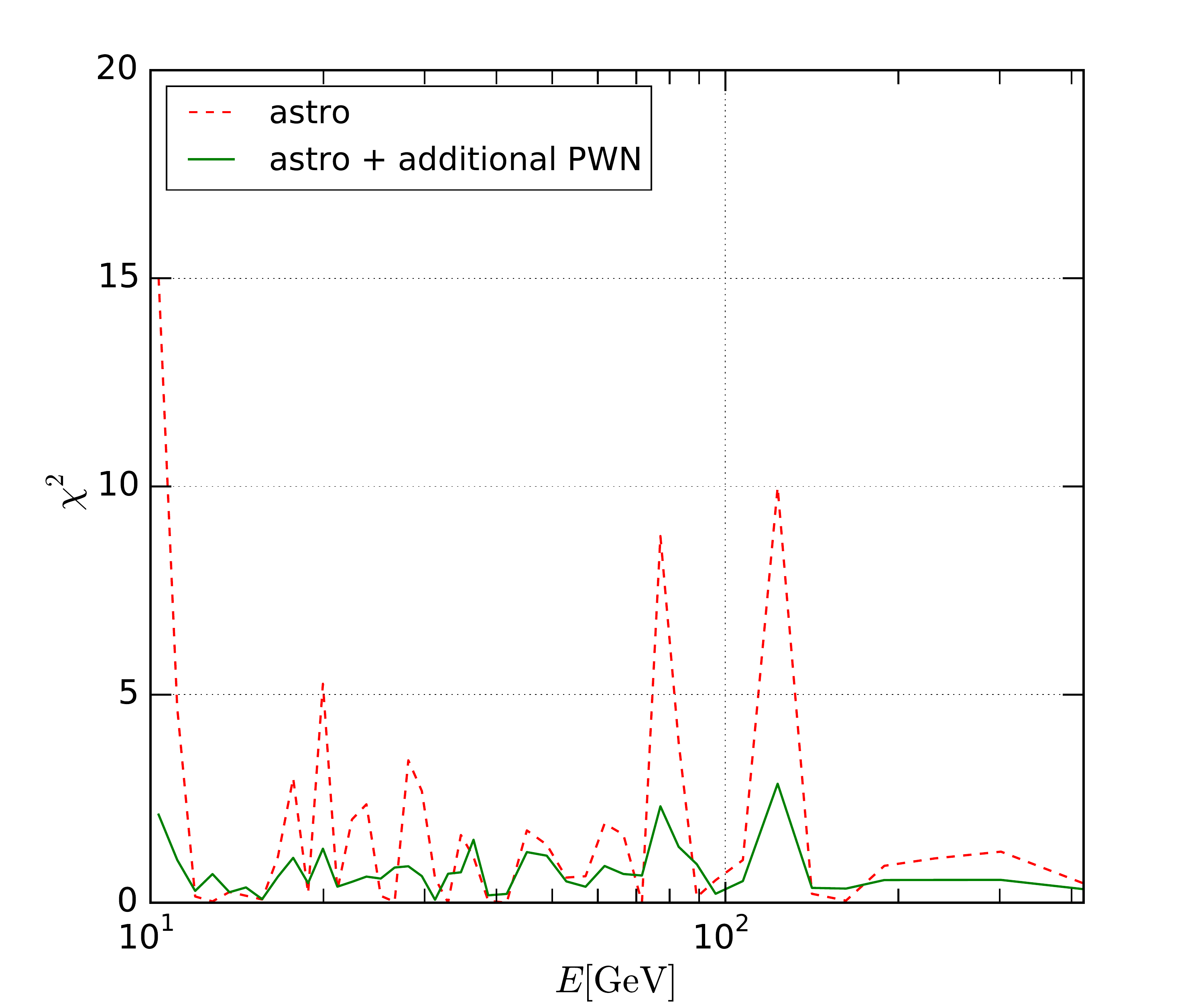}
\caption{{\it Left panel}: the positron fraction for the best-fit configurations of the {\it astro} model and of the {\it astro + additional PWN} model are shown together with AMS-02 data as a function of the energy. {\it Right panel}: the $\chi^2$ associated to the two models is shown as a function of the energy.}
 \label{fig:psr_chi}
 \end{figure}
 
\begin{figure}
\center
\includegraphics[width=0.47 \textwidth]{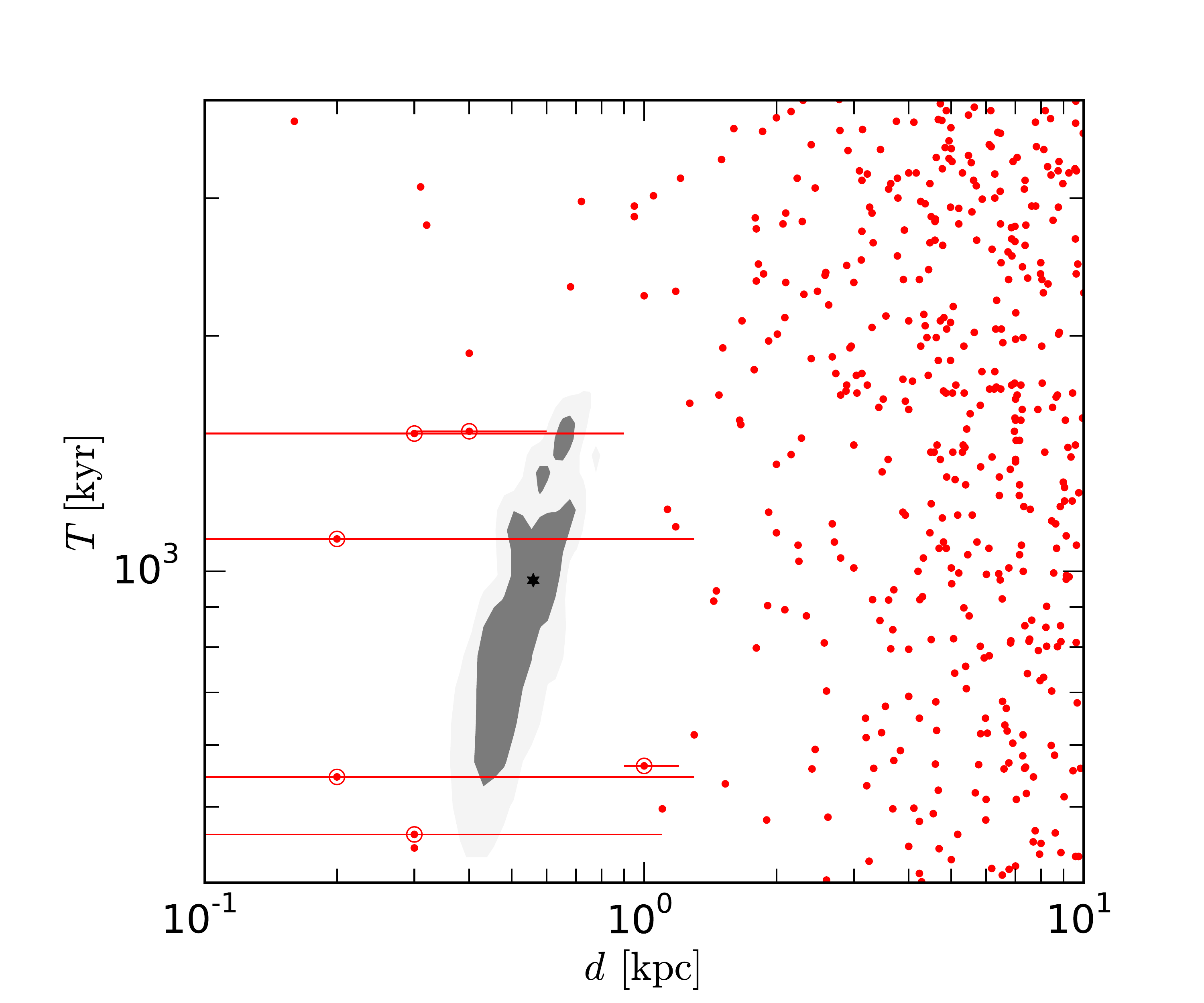}
\caption{Contour plot of the distance and age of the additional pulsar together with the other sources from the ATNF catalogue; the star denotes the best-fit, the darker and lighter shadings represent, respectively, the $1\sigma$ and $2\sigma$ confidence regions, while the bars quantify the $1\sigma$ uncertainties on the position of some of the ATNF catalog pulsars (these sources are here circled in red, for clarity).}
 \label{fig:psr}
 \end{figure}

The result of the fit is shown in Table \ref{tab:psr}. The left panel of Fig.~\ref{fig:psr_chi} shows the predicted positron fraction for the best-fit configuration compared to the prediction given by the {\it astro} model, together with the AMS-02 data, while the right panel illustrates the behaviour of the $\chi^2$ as a function of the energy for the two models. The ``fictitious'' pulsar that we are adding sizably improves the agreement with the AMS-02 data as compared to the {\it astro model} and provides a fit that is comparable to the ones that are associated to a DM signal in the muon channel, as reported in Table \ref{tab:bestfits}. For the positron fraction dataset, the agreement is extremely good: the chi-square per data point is 0.93. This occurs if this source is able to provide a non-negligible contribution in the low/intermediate-energy region of the positron flux, which, as already stressed, is where also the DM contribution from leptonic annihilation/decay channels was preferred by the fit. The best-fit configuration and the $1\sigma$ and $2\sigma$ C.L. regions for the additional pulsar in the $(d_{\rm psr},T_{\rm psr})$ plane are reported, together with the PWNe listed in the ATNF catalogue, in Fig. \ref{fig:psr}. The red bars denote the 1 $\sigma$ uncertainty associated to the position of some of the sources that are among the closest to the contour regions of the model (as reported in \cite{2012ApJ...755...39V}). As one can see, in order to have a seizable flux at these energies, the pulsar should be relatively close (600 pc), young (1000 kyr) and powerful ($\eta_{\rm{psr}} W_0 = 1.4^{+0.2}_{-0.4} \times10^{39}$ erg). Moreover, the solution appears to fall in a region where the ATNF catalogue has no sources. However, once that the uncertainty on the determination of the distance of the pulsars is taken into account, some of the sources of the catalogue may overlap with the contour regions of the best-fit that we find. In addition, one must always remember that this model could be refined by allowing for a free spectral index of the fictitious source. This could impact the reconstructed values of the age and distance of the putative source. With all this considered, we cannot exclude that this model could represent a perfectly  viable interpretation of AMS-02 data.

\
 \begin{table}[t]
\centering
\footnotesize
\begin{tabular}{|c|c|c|c|c|c|c|}
\hline
\multicolumn{5}{|c|}{{\bf Annihilating DM}} \\
\hline
Parameter & astro& $e^+ e^-$ & $\mu^+\mu^-$ & $\tau^+ \tau^-$ \\
\hline
$\eta_{PWNe}$ & $ 19.7\substack{+90.5 \\ -9.7} \times 10^{-4} $ & $ 18.27\substack{+94.25 \\ -17.25} \times 10^{-4} $ & $ 31.7\substack{+112.5 \\ -30.7} \times 10^{-4} $ & $ 3.5\substack{+98.7 \\ -2.5} \times 10^{-4} $ \\ 
\hline
$\gamma_{PWNe}$ & $1.64 \substack {+ 0.76 \\ -0.70}$ & $1.70 \substack {+ 0.50 \\ -0.25}$  & $1.69 \substack {+ 0.51 \\ -0.24}$ & $1.67 \substack {+ 0.53 \\ -0.22}$ \\ 
\hline
$\eta_{1}$ & $0.104 \substack {+0.035\\-0.071}$  & $0.089 \substack {+0.041\\-0.080}$ & $0.054 \substack {+0.064\\-0.053}$  & $0.111 \substack {+0.030\\-0.101}$\\ 
\hline
$\eta_{2}$ & $0.005 \substack {+0.027\\-0.004}$  & $0.011 \substack {+0.032\\-0.010}$  &  $0.001 \substack {+0.077\\-0.001}$ & $0.071 \substack {+0.035\\-0.007}$ \\ 
\hline
$\eta_{3}$ & $0.023 \substack {+0.066\\-0.022}$ & $0.005 \substack {+0.089\\-0.005}$  & $0.023 \substack {+0.068\\-0.022}$ & $0.013 \substack {+0.107\\-0.013}$ \\ 
\hline
$\eta_{4}$ & $0.297 \substack {+0.056\\-0.131}$  & $0.299 \substack {+0.063\\-0.137}$  & $0.338 \substack {+0.096\\-0.171}$ & $0.290 \substack {+0.107\\-0.154}$ \\ 
\hline
$\eta_{5}$ & $0.006 \substack {+0.185\\-0.005}$  & $0.014 \substack {+0.206\\-0.014}$  & $0.027 \substack {+0.168\\-0.026}$  & $0.017 \substack {+0.196\\-0.001}$  \\ 
\hline
$Q_{0, SNRs} [10^{50}$ erg/s] & $1.12 \substack {+ 0.17\\ -0.05}$ & $1.08 \substack {+ 0.18\\ -0.03}$ & $1.10 \substack {+ 0.10\\ -0.05}$ & $1.14 \substack {+ 0.18\\ -0.06}$  \\ 
\hline
$\gamma_{SNRs}$ & $2.22\substack{+0.03\\-0.01}$ & $2.21\substack{+0.03\\-0.01}$ & $2.22\substack{+0.01\\-0.02}$ & $2.22\substack{+0.03\\-0.01}$ \\ 
\hline
$N_{Vela}$ & $0.79\substack{+0.19\\-0.21}$  & $0.89\substack{+0.41\\-0.04}$  & $0.65\substack{+0.25\\-0.11}$  & $0.79\substack{+0.18\\-0.23}$ \\ 
\hline
$m_{DM}$ [GeV] & - & $34\substack{+130\\-14}$ & $82 \substack{+108.\\-44.}$  & $142\substack{+35\\-9}$ \\    
\hline
$\langle \sigma v \rangle $ [cm$^3$s$^{-1}$]& - &$1.34\substack {+14.3\\-1.2}\times 10^{-27}$ & $4.2 \substack {+4.3\\-4.2}\times 10^{-26}$  & $2.6 \substack{+20.2\\-2.5}\times 10^{-26}$ \\ 
\hline
$\chi^2/ {\mathrm {d.o.f.}}$ & 1.00 & 0.97  & 0.93 & 0.91 \\ 
\hline
%$p$-value(sign.) & $2.2\times10^{-2}(2.2\sigma)$ & $1.2\times10^{-2}(2.5\sigma)$ & $2.7\times10^{-4}%(3.6\sigma)$ & $4.8\times10^{-3}(2.8\sigma)$ \\
\end{tabular}
\caption{Best-fit configurations for the {\it refined astro model}.}
\label{tab:5psrs}
\end{table}

As a second extension of the analysis of the previous Sections, and in order to investigate at a deeper level what we can do with the sources that are in the ATNF catalogue,  we build a more complex astrophysical framework, which we label {\it refined astro model}, that is characterized by the assumption that the whole set of PWNe listed in the ATNF catalogue do not share anymore the same efficiency $\eta_{\mathrm{PWNe}}$. Instead, we single out the five most powerful sources of the catalogue in the different energy ranges by using the same ranking algorithm that has been used in Ref.~\cite{DiMauro:2014iia} and we associate to their emissions the efficiencies $\eta_i$ (being $i=1,\dots,5$) that act as free parameters in our fitting procedure. These most powerful sources are \cite{DiMauro:2014iia}: Geminga ($i=1$), J2043+2740 ($i=2$), J0538+2817 ($i=3$), Monogem ($i=4$) and B1742-30 ($i=5$). To limit the number of parameters in the fit we associate to each one of these five sources the same spectral index $\gamma_{i=1,\dots,5} = 1.8$ \cite{DiMauro:2014iia} (we have checked that different values for this parameter does not affect our conclusions).  The parameters $\eta_{\rm PWNe}$ and $\gamma_{\rm PWNe}$, whose definition has been given in the previous Sections, are still present and characterize the population of the {\it other} PWNe in the ATNF catalogues, which behave like a background of sources that share common values for their efficiencies and spectral indices. To summarize: we consider emission from the full ATNF catalogue, but for the 5 most powerful sources we allow for a free normalization parameter on their emission.

In Table~\ref{tab:5psrs} we list the best-fit configurations that are obtained by fitting AMS-02 data within the {\it refined astro model}, {\it without} and {\it with} the addition of DM. We illustrate only the results obtained for the annihilating case, since we do not register any significant change for a DM that decays in the same channels (except for the fact that the best-fit DM masses are halved). We show only the best-fit configurations for the leptonic channels, since for DM annihilation into the $b\bar{b}$ and $W^+W^-$ channels the addition of a DM signal does not improve the fit as compared to the purely astrophysical case. 

\begin{figure}[t]
\centering
\includegraphics[width=0.69 \textwidth]{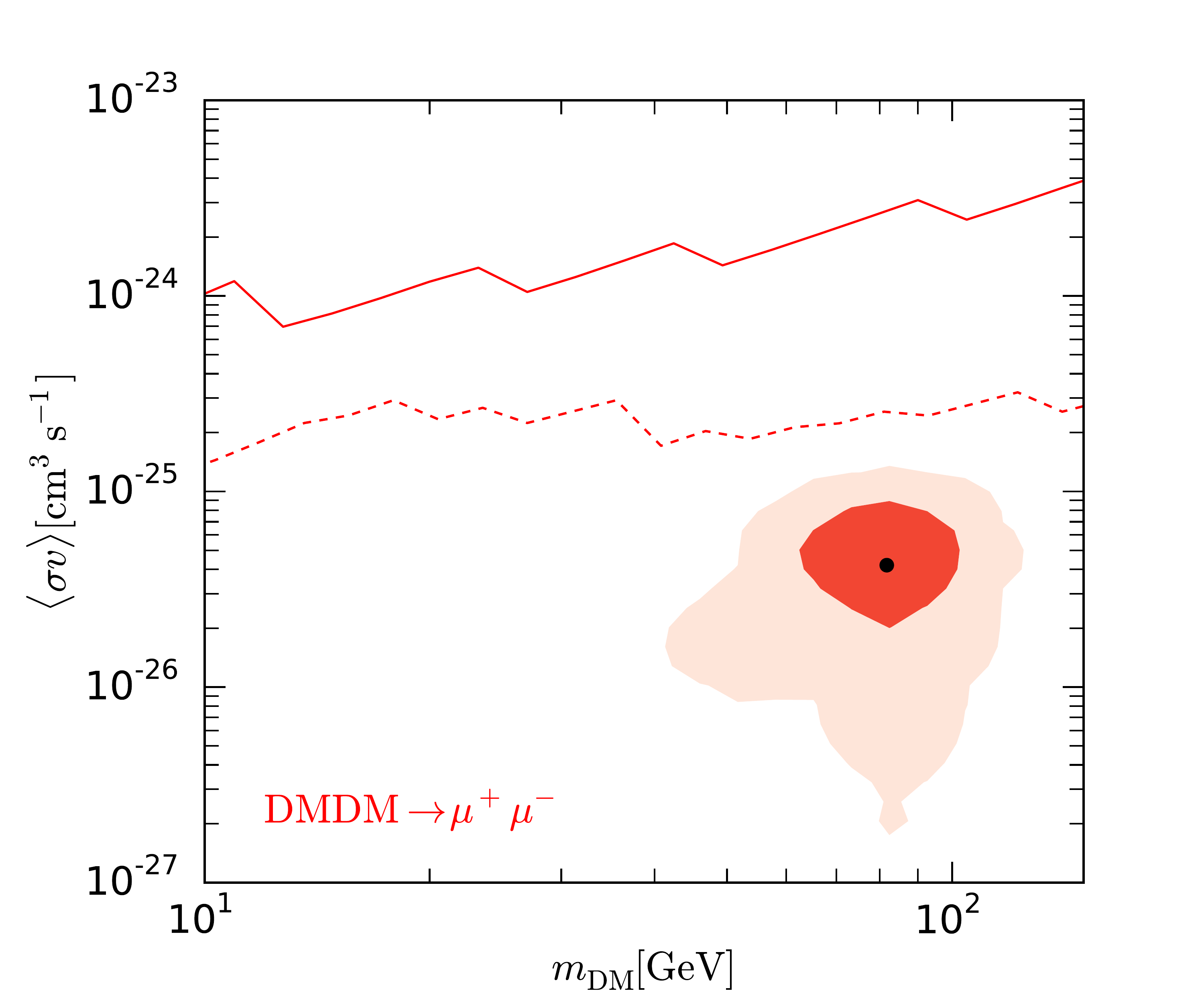}
\caption{$1\sigma$ and $2 \sigma$ allowed regions in the $(m_{DM},\sigmav)$ plane for a DM annihilating in the $\mu^+\mu^-$ channel in the {\it refined astro model with DM}. Solid and dashed lines represent, respectively, the conservative and optimistic gamma-ray upper limits derived in Refs.~\cite{Calore:2013yia,DiMauro:2015tfa}.}
\label{fig:CP3}
\end{figure} 

\begin{figure}[t]
\centering
\includegraphics[width=0.49 \textwidth]{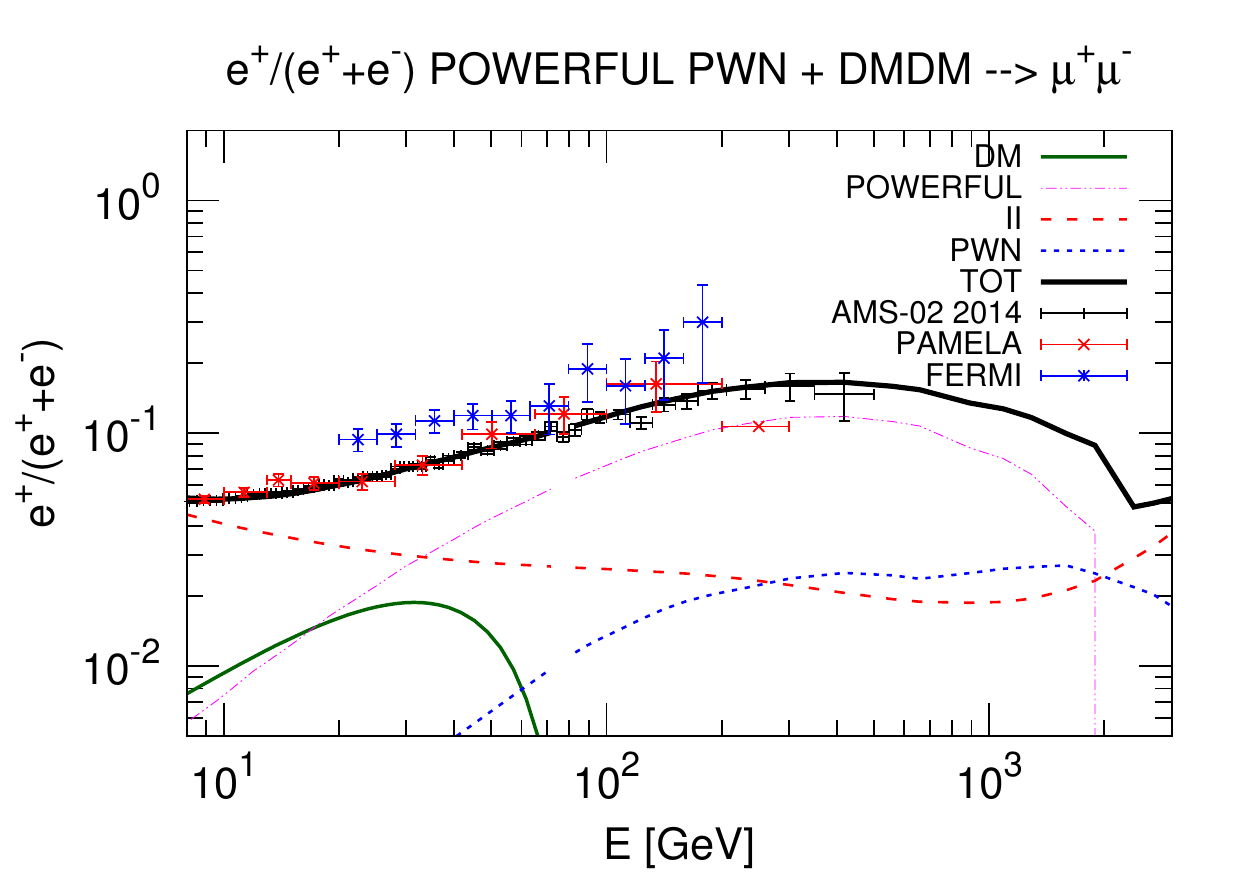}
\caption{The contribution to the positron fraction associated to the best-fit configuration for a DM annihilating in $\mu^+ \mu^-$ channel (solid green line) shown together with the flux from the five most powerful sources (dot-dashed magenta line), from all the other pulsars of the ATNF catalog (dotted blue line) and from secondary production (dashed red line)}
\label{fig:CP2}
\end{figure}

In the case of a pure astrophysical interpretation, the results in terms of the reconstructed parameters are stable as compared to the ones obtained in the {\it astro model}, with the notable exception of the emission coefficient $\eta_i$ for the 5 most powerful PWN: Geminga and Monogem are required to be the dominant emitters. The catalogue-PWN other than the 5 most powerful are now significantly suppressed as compared to the previous analysis of the {\it astro model}: $\eta_{PWNe}$ drops from 3.7\% to 0.1\%. This implies that the AMS-02 data could be shaped by very few dominant sources, like Monogem and Geminga. The significance of the fit is here improved as compared to the simpler {\it astro model} and makes this model perfectly compatible with AMS-02 data. This holds also for the positron fraction dataset for which the chi-square per data point is 1.02.

In the case a DM signal is present, we see that within this more complex model for the astrophysical emission, the contribution from a DM annihilating into leptons is able to improve only very slightly the fit to AMS-02. The chi-square per data point of the positron fraction is reduced only for the $\mu^+ \mu^-$ annihilation channel, and reaches 0.9. 
%in particular, for the annihilation into $\mu^+ \mu^-$ and $\tau^{+}\tau^{-}$, the significance is definitely higher than in the {\it astro+DM model}: the $p$-values of the best fits improve from $1.1\times 10^{-2}$ (significance of $2.5\sigma$) to $2.7\times 10^{-4}$ ($3.6\sigma$) for the $\mu^+ \mu^-$ channel and from $5.8\times 10^{-2}$ ($1.9\sigma$) to $4.8\times 10^{-3}$ ($2.8\sigma$) for the $\tau^{+}\tau^{-}$ channel. 
Concerning gamma-ray upper limits, the best-fit configuration for the $\tau^+ \tau^-$ case is still in tension with the constraints shown in Fig.~\ref{fig:CP}, while the $\mu^+ \mu^-$ DM candidate is fully compatible with these bounds, and therefore represent the best option for a DM interpretation. In the case of the preferred $\mu^+ \mu^-$ channel, the best-fit mass in the {\it refined astro model} turns out to be almost the same as in the {\it astro model} (82 GeV instead of 89 GeV)  and the annihilation cross section is almost a factor of two smaller, being compatible with the thermal value at the $2\sigma$ level. The allowed region and the gamma-rays bounds for the  $\mu^+ \mu^-$ channel are shown in Fig.~\ref{fig:CP3}. 
%We wish to stress that the $\Delta \chi^2$, namely the difference of the chi-square considering the refined purely-astrophysical model and the model with the addition of DM annihilating into $\mu^+\mu^-$ is 24, associated to a significance of the preference of DM in our model of about $4.5\sigma$. Therefore in the contest of our astrophysical model the DM contribution with a $\mu^+\mu^-$ annihilation channel is significantly preferred and is fully consistent with $\gamma$-ray bounds, and with constraints arising from the CMB \cite{Slatyer:2015jla,Galli:2011rz}.
We notice also that in this case, Monogem is the largely dominant pulsar emitter.

The positron fraction corresponding to the best fit results for the {\it refined astro model with DM}, which occurs for the $\mu^+ \mu^-$ channel, is shown in Fig.~\ref{fig:CP2}. Notice that in this case, where a single pulsar largely dominates the positron flux (Monogem), the positron fraction is predicted to drop relatively quickly at the pulsars cut-off energy (which in our analysis is taken at 2 TeV). This feature is here due not to the DM emission (which, in this case, gives its maximal contribution at intermediate energies), but to the pulsars high energy properties. Data above the TeV scale could help in distinguishing the situation where all known pulsars contribute in a democratic way ({\it astro model}, with or without DM), from a case where a single pulsar dominates the high-energy positron emission (in our analysis, the {\it refined astro model}, again with or without a DM contribution).

\section {Conclusions}  
\label{sec:conclusions}

In this paper we have performed a quantitative study of AMS-02 electron and positron data, in terms of both galactic known sources and a possible 
dark matter contribution. 

Firstly, by updating the analysis illustrated in Ref.~\cite{DiMauro:2014iia}, we have investigated the possibility to interpret the whole set of leptonic AMS-02 data in terms of astrophysical sources of primary and secondary origin only. We have shown that a model based on the simplifying assumptions that all the PWNe of the ATNF catalogue share the same efficiency and spectral index is in {\blue} some tension with the measured positron fraction. 

Secondly, we have carried out a detailed study of the interplay between the contributions to the positron flux that derive from primary astrophysical sources (namely PWNe) and from the annihilation/decay of DM particles. 
We have worked within a scenario in which DM gives its contribution on top of a realistic astrophysical leptonic emission, which acts 
as a non trivial background with respect to a putative DM signal. 
On one hand, we have investigated how one can use the highly precise data provided by the AMS-02 experiment to derive robust constraints to DM properties. On the other hand, we have illustrated how the addition of a DM contribution compatible with the bounds that arise from other indirect detection channels, can improve the fit to AMS-02 data as compared to the simple {\it astro} model. Specifically, we have found that a DM particle with a mass around 80 GeV, annihilating in the $\mu^+\mu^-$ channel with a cross section  remarkably close to the thermal value can provide, when added to the PWNe  and to the secondary contributions, an excellent fit to the AMS-02 data, positron fraction included. Conversely, the addition of a DM annihilating or decaying into hadronic channels, does not appear to improve the fit significantly and the best-fit configurations for these channels are in tension with upper bounds derived from different indirect DM searches (e.g. gamma-rays). 

We have also investigated whether the addition of a local pulsar to the sources of the ATNF catalogue can provide a fit to the AMS-02 data of comparable goodness as the DM hypothesis. This is indeed possible and, within this theoretical framework, one is able to obtain a remarkably good fit to AMS data. The characteristics (distance and age) of this additional pulsar do not correspond exactly to any source already present in the ATNF catalogue. However, once that the uncertainty associated to the distance of these sources is taken into account, some of them appear to possess characteristics that are compatible with the best-fit region of the model. This makes this interpretation of AMS-02 data perfectly viable.

Finally, we have shown that relaxing the hypothesis of equal efficiency and spectral index for all the PWNe of the ATNF catalogue
weakens the need to have additional sources, and a pure astrophysical interpretation of AMS-02 data is indeed possible.

We conclude by saying that the parallel inspection of independent and multi-wavelength channels is a crucial step in pursuing a deeper understanding of the astrophysical sources that populate our Galactic environment. This investigation, together with the precise measurements performed by current and future experiments may significantly help in interpreting the cosmic lepton data and, perhaps, in shedding light on the DM mystery.

%%%%%%%%%%%%%%%%%
\acknowledgments

This work is supported by the research grant {\sl Theoretical Astroparticle Physics} number 2012CPPYP7, funded under the program PRIN 2012 of the Ministero dell'Istruzione, Universit\`a e della Ricerca (MIUR), by the research grant {\sl TAsP (Theoretical Astroparticle Physics)} funded by the Istituto Nazionale di Fisica Nucleare (INFN), and by the {\sl Strategic Research Grant: Origin and Detection of Galactic and ExtraGalactic Cosmic Rays} funded by Torino University and Compagnia di San Paolo. A.V acknowledges the hospitality of the IPhT CEA Saclay and of the Institut d'Astrophysique de Paris (IAP) where part of this work was done.

%%%%%%%%%%%%%%%%%

\appendix
\section{Fitting different datasets}
\label{app:datasets}

\begin{figure}[t]
\centering
\includegraphics[width=0.44 \textwidth]{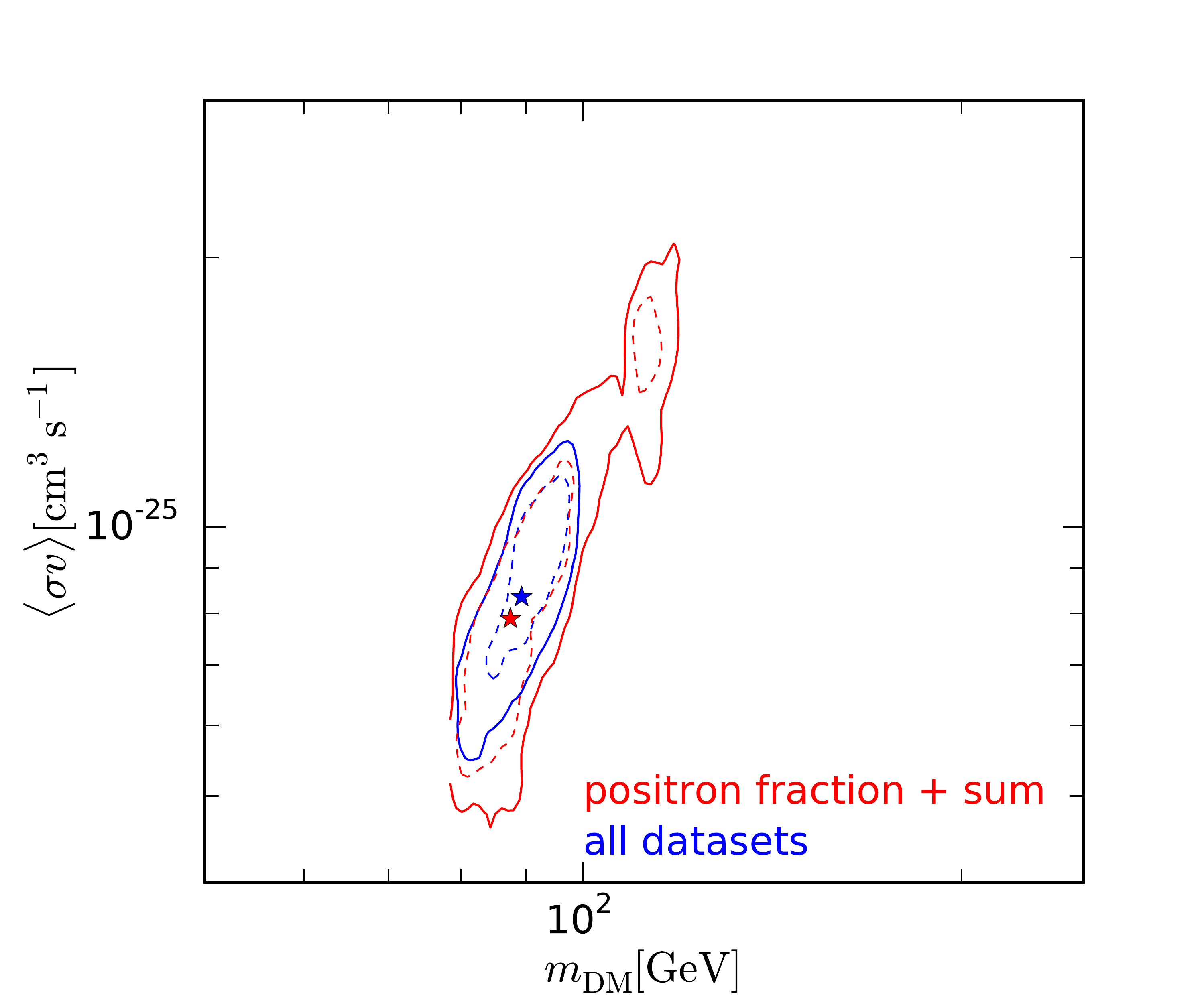}
\includegraphics[width=0.44 \textwidth]{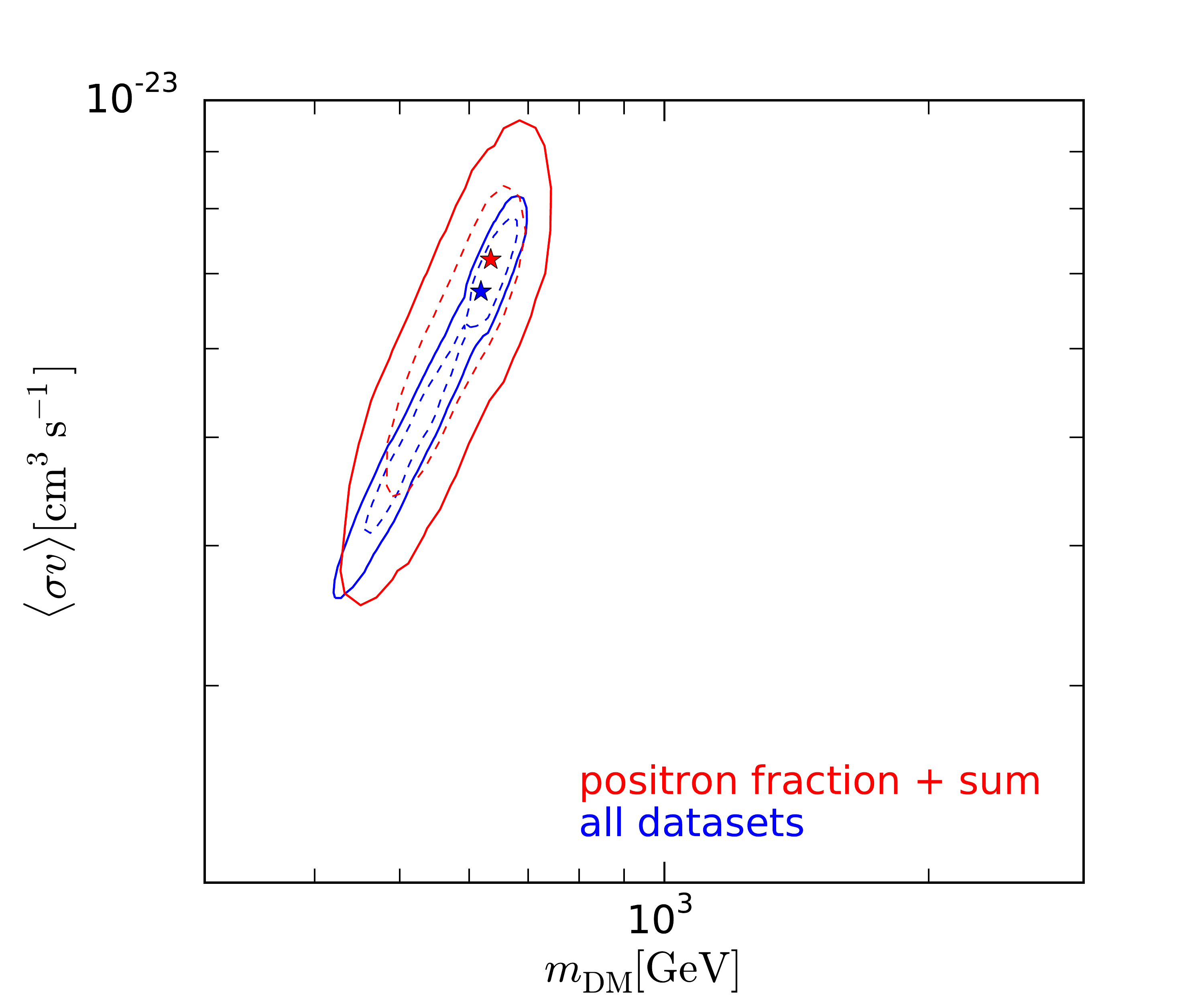}\\
\includegraphics[width=0.44 \textwidth]{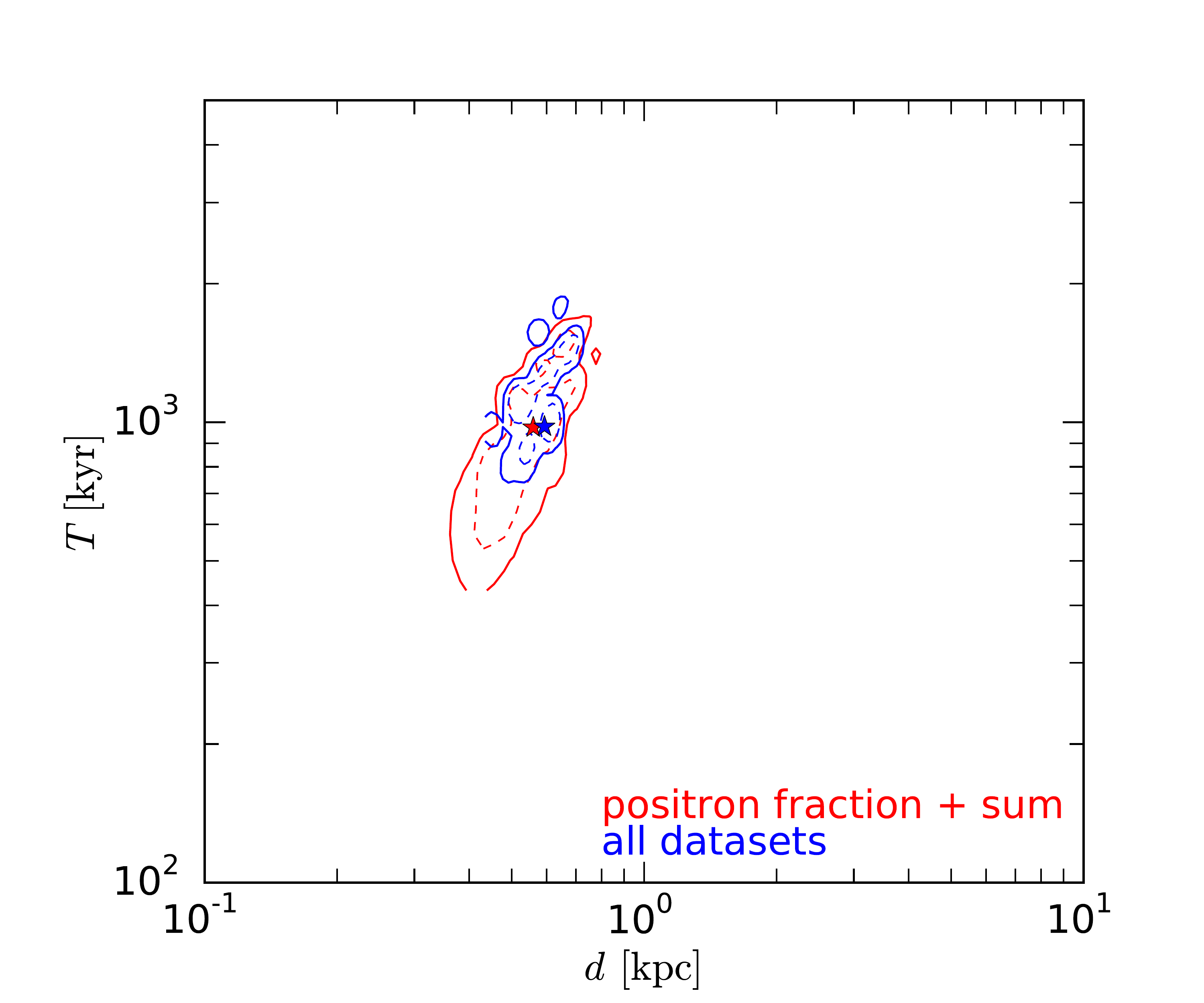}
\caption{$1 \sigma$ and $2 \sigma$ contour regions 
for different models and different choices of the datasets included in the fit.
Blue lines refer to the fit of all the four AMS-02 datasets,
while red lines refer to the fit to the positron fraction and total
flux only. Stars denote the best-fit configurations. The contour regions
reported in the top row refer to the astro + DM model for two different
annihilation channels: $\mu^+\mu^-$ in the left panel and $\tau^+\tau^-$ in the right panel. The
bottom row shows the contours in the $(d,T)$ plane for the additional PWN
discussed in Section \ref{sec:psr}. }
\label{fig:datasets}
\end{figure} 

\begin{figure}[t]
\centering
\includegraphics[width=0.49 \textwidth]{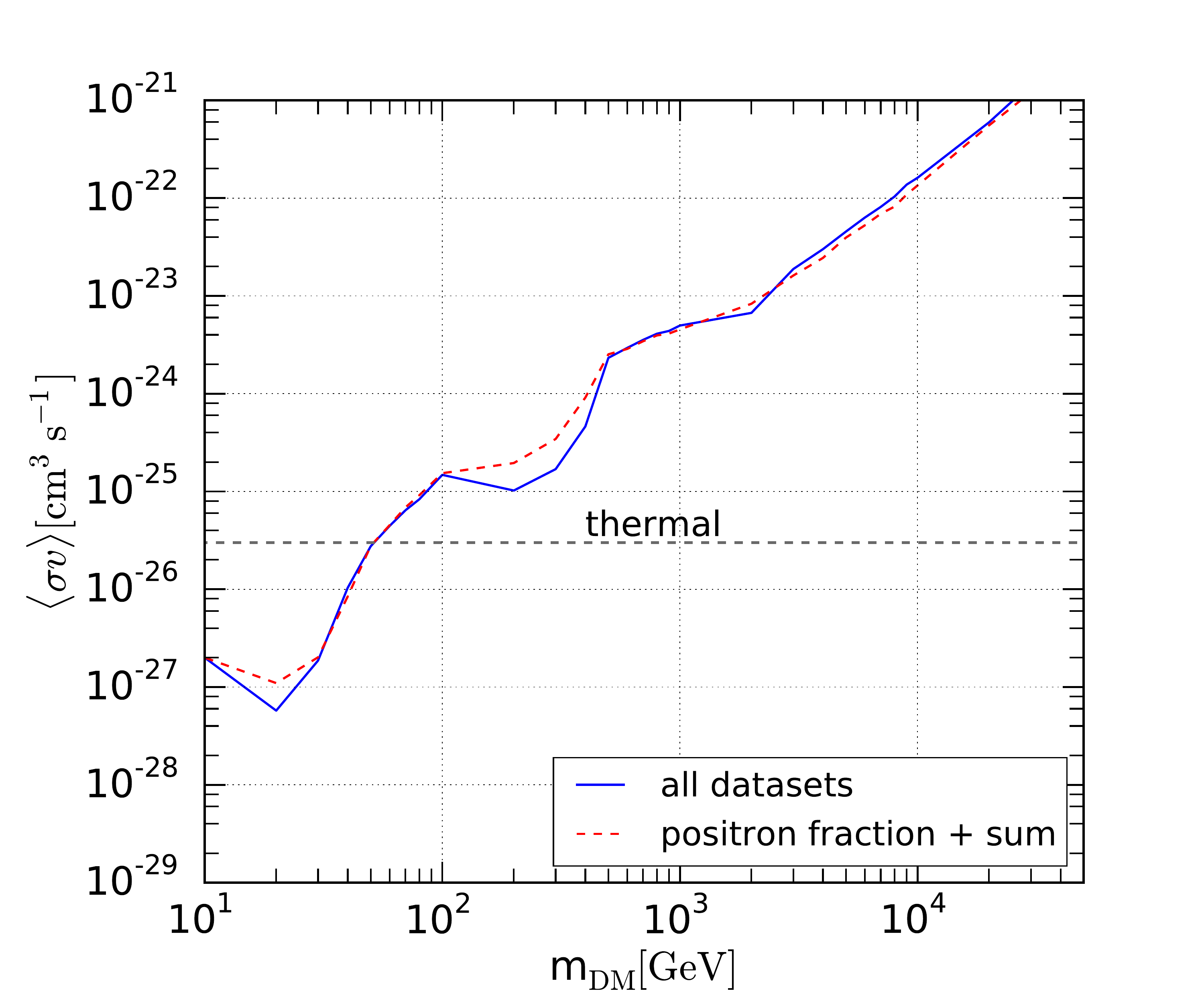}
\caption{Upper limits to the DM annihilation cross section, for the $\mu^+\mu^-$ annihilation channel, for different choices of datasets included in the fit, as reported in the boxed inset. }
\label{fig:UL_datasets}
\end{figure} 

As discussed in Section \ref{sec:results}, all the results shown in this paper have been derived by fitting only the {\it positron fraction} and {\it total flux} datasets. In fact, as we have already discussed, since the four observables measured by AMS-02 are not fully independent, including additional datasets in the fit could potentially lead to an overfit of the data and bias our results. \\
To illustrate this point, we plot in Fig. \ref{fig:datasets} the 1$\sigma$ and 2$\sigma$ contour regions for the best-fit parameters of different models obtained by fitting different sets of data (we show the {\it astro + DM} model in the top row and the {\it astro} model with an additional PWN in the bottom row). In particular, the blue contours are the ones obtained by fitting the four datasets, while the red ones are the result of fitting only the positron fraction and the total flux. It is manifest that fitting all the four datasets determines a shrinking of the contour regions. The best-fit values of the parameters of the different models that we consider seem to be not significantly affected by the choice of the datasets to include in the fit, but the value itself of the chi-square changes notably. In fact, when fitting the four datasets, the best fit configurations are associated to $\chi^2/{\mathrm {d.o.f.}} = 0.78$ ({\it astro + DM} model, $\mu^+\mu^-$ channel),  $\chi^2/{\mathrm {d.o.f.}} = 0.84$ ({\it astro + DM} model, $\tau^+\tau^-$ channel), $\chi^2/{\mathrm {d.o.f.}} = 0.70$ ({\it astro} model + additional PWN) while, for the same models, we have $\chi^2/{\mathrm {d.o.f.}} = 0.98$, $\chi^2/{\mathrm {d.o.f.}} = 1.05$ and $\chi^2/{\mathrm {d.o.f.}} = 0.85$ when fitting only the positron fraction and the total flux. Usually, values of the reduced chi-square that are significantly less than 1 represent a clear indication that the choice of datasets is such that the model is overfitting data.
This is the reason why, for all the analyses shown in the paper, we have considered fits to the positron fraction and the total flux only. \\
Lastly, in Fig. \ref{fig:UL_datasets}, we show how the choice of the datasets to be included in the fit affects the determination of the upper limit to the DM annihilation cross section for the $\mu^+\mu^-$ annihilation channel. As one could have easily expected, in connection with the shrinking of the contour regions discussed in the previous paragraph, for the parameters  around the best-fit ones, the bounds obtained with fits to only the positron fraction and the total flux are weaker than the ones obtained by fitting the four datasets. On the other hand, for all the masses for which the addition of DM does not improve the quality of the fit with respect to the {\it astro model}, bounds are rather insensitive to the choice of dataset.

%%%%%%%%%%%%%%%%%

\clearpage
\bibliography{positronsDM}

 %%%%%%%%%%%%%%

%%%%%%%%%%%%%%

%%%

\end{document}